\documentclass[aps,pre,twocolumn,notitlepage,showpacs,superscriptaddress,groupedaddress]{revtex4-1}  

\usepackage{graphicx,epstopdf}  
\usepackage{amsmath}
\usepackage{graphicx}
\usepackage{amsfonts}
\usepackage{color}
\usepackage{dcolumn}   
\usepackage{bm}        
\usepackage{amssymb}   
\usepackage{hyperref}
\begin{document}

\title{Equilibrium dynamical correlations in the Toda chain and other integrable models}%

\author{Aritra ~Kundu and Abhishek ~Dhar}
 \affiliation{International Centre for Theoretical Sciences (TIFR), Survey No.~151, Shivakote,~Hesaraghatta Hobli,~ Bengaluru~ -~ 560 089,~ India.}
 \email{aritrak@icts.res.in,abhishek.dhar@icts.res.in }
 
\date{\today}

\newcommand{\angstrom}{\text{\normalfont\AA}}
\newcommand{\braket}[3]{\bra{#1}\;#2\;\ket{#3}}
\newcommand{\projop}[2]{ \ket{#1}\bra{#2}}
\newcommand{\ket}[1]{ |\;#1\;\rangle}
\newcommand{\bra}[1]{ \langle\;#1\;|}
\newcommand{\iprod}[2]{\bra{#1}\ket{#2}}
\newcommand{\logt}[1]{\log_2\left(#1\right)}
\def\cI{\mathcal{I}}
\newcommand{\cx}[1]{\tilde{#1}}
\newcommand{\nn}{\nonumber}
\newcommand{\la}{\langle}
\newcommand{\ra}{\rangle}
\newcommand{\p}{\partial}
\def\be{\begin{equation}}
\def\ee{\end{equation}}
\def\bea{\begin{eqnarray}}
\def\eea{\end{eqnarray}}

\newcommand{\eqa}[1]{\begin{align}#1\end{align}}
\newcommand{\mbf}[1]{\mathbf{#1}}
\newcommand{\iu}{{i\mkern1mu}}

\begin{abstract}
We investigate the form of equilibrium spatio-temporal correlation functions of conserved quantities in the Toda lattice and in other integrable models. From numerical simulations we find that the correlations satisfy ballistic scaling with a remarkable collapse of data from different times. 
We examine special limiting choices of parameter values, for which the Toda lattice tends to either the harmonic chain or the equal mass hard-particle gas. In  both these limiting cases, one can obtain the correlations exactly and we find excellent agreement with the direct Toda simulation results. We also discuss a 
transformation to ``normal mode'' variables, as commonly done in hydrodynamic theory of non-integrable systems, and find that this is useful, to some extent, even for the integrable system. The striking differences between the Toda chain and a truncated version, expected to be non-integrable, are pointed out.  
\end{abstract}

\maketitle

\section{Introduction}

There has been a lot of recent interest in  equilibrium correlations of conserved quantities in one-dimensional Hamiltonian systems, in particular in the form of the temporal-relaxation for a system in equilibrium. Remarkable predictions have been obtained for the form of spatio-temporal correlations in systems of one-dimensional fluids and anharmonic chains, using the framework of 
fluctuating hydrodynamics \cite{HenkPRL2012,MendlSpohnPRL13,SpohnJStat14,RamaswamyPRL02}. 
For generic nonlinear systems with three conserved quantities (mass, momentum, energy), it has been predicted that 
there are two sound modes which exhibit correlations as those in the Kardar-Parisi-Zhang (KPZ) equation, and a single heat mode showing characteristics of a Levy walk. These predictions have been verified in many systems \cite{mendl14,DasDharPRE14,spohn15}.  
These studies of equilibrium fluctuations of conserved quantities have led to some progress in resolving the long standing puzzle of anomalous heat transport in one dimensional systems \cite{Lepri0PR3,DharAdvPhy08,Book}. The general consensus from about two decades of theoretical and numerical studies is that in one dimensional momentum conserving non-integrable systems, the heat transport is anomalous, that is the heat conductivity ($\kappa$) diverges with system size ($\kappa \sim N^{\alpha}$, where $0 \leq \alpha \leq 1 $ ). The decay of equilibrium fluctuations shows similar anomalous features which lead to a understanding of the nonequilibrium state via linear response.

An important aspect, which affects transport and fluctuations in a many-body system,  is  the integrability of the Hamiltonian.  It is widely believed that, if we apply different temperatures to the two ends  of an integrable system, then the energy current would not decay with system size (for large systems). This is referred to as ballistic transport. It is expected in non-integrable systems that typical non-linear interactions should lead to sufficient effective stochasticity in the dynamics, which should then cause a decay of the heat current with system size. Similarly the decay of equilibrium fluctuations is expected to be ballistic, i.e, the width of the correlation function spreads in time as $\sim t$. There are very few simulations exploring equilibrium correlations of conserved quantities in integrable models \cite{TheodorPRL99,Zhao2006,ShastryYoungPRB10}.  
Also we have not found a mathematical statement in the literature of the conditions under which one gets  non-decay, with system-size,  of the energy current in the open-system with applied temperature bias, and of ballistic scaling of correlations functions. 
In fact there are examples of integrable systems \cite{AdharPRL99}, in one dimensions with stochastic dynamics and in higher dimensions with Hamiltonian dynamics, where one has diffusive transport (though no local equilibrium). 
Some exact results are known on properties of the equilibrium energy current correlation function, for example in terms of Mazur  inequalities \cite{ZotosJLTP02}. But it is not clear what this precisely means for either the decay of equilibrium spatio-temporal correlations, or for the system size dependence of current in the non-equilibrium setting.

The main aim of this paper is to perform a detailed study of the form of equilibrium correlations in integrable systems.
In particular we study the well known model of an anharmonic chain, the Toda chain  \cite{TodaSpringer12}, first introduced in 1967  as  example of an integrable one-dimensional (1D)  system which  generalizes the harmonic chain to large amplitude oscillations. The chain is characterized by non-linear interactions of exponential type between nearest neighbors while still being integrable. The exact solvability of the model was studied in \cite{FlaschkaPhysRev74,HenonPhysRev74} 
where it was reported how to construct a full set of  conserved quantities using the 
Lax pair formalism. The periodic lattice was studied in \cite{KacMorPNAS75,TanakaPTPS76} using the inverse scattering method. In the limit 
of large anharmonicity the chain is characterized by soliton solutions, which are stable wave packets localized in real space. For infinite chains the isolated soliton solution was found by Toda \cite{TodaSpringer12}. For periodic finite chains one can find exact solutions, the so-called  cnoidal waves, which are periodic trains of solitons  \cite{TodaSpringer12,ShastryYoungPRB10}. The equilibrium thermodynamic properties such as specific heat, etc \cite{TodaSaitoh} can be studied by performing exact integrals with respect to Gibbs distribution. Although special exact classes of solutions are known for the Toda chain, finite temperature dynamical properties such as correlation functions are hard to access analytically. There have been some attempts to study finite temperature dynamical properties of Toda chain through non-interacting soliton gas analogy \cite{ButtnerMertnesSSC79, BauerMertensPhyB88, TheodorPRL99} and through taking classical limit of a quantum Toda chain \cite{CuccoliPRB93}. 
The quantum Toda chain was solved in \cite{Gutzwiller80} and by the Bethe ansatz  in \cite{SutherlandRMJM78}. 
A review of various static and dynamic properties of the Toda chain can be found in  \cite{CuccoliIJMPB94}.

Energy transport in the Toda chain was studied in \cite{ZotosJLTP02}, where the decay of current correlations and  overlap of currents with other conserved quantities  were studied in the context of Mazur inequalities. 
A careful numerical study was carried out in \cite{ShastryYoungPRB10} looking at the decay of current correlations in finite systems prepared in canonical equilibrium.  
It was pointed out that the Mazur relations needed to be modified and that one needed to take projections of the current to not just the conserved quantities but also to their bilinear combinations.
Among the other results in \cite{ShastryYoungPRB10}, the existence of special ``cnoidal" solutions in the periodic Toda chain was noted and the effect of cubic and quartic perturbations on the decay of conserved quantities was studied. 

To test the role of integrability in heat transport, it is interesting to study transport in perturbed integrable systems. The effect of solitons on the heat transport in Toda chain and its perturbations was studied in \cite{TodaPhyScript79}. The diatomic alternate mass Toda chain which is non-integrable was studied in \cite{HatanoPRE99} where it was found that the thermal conductivity $\kappa$ diverges with system size $N$ as as $\kappa \sim N^{0.34}$. 
Heat transport in Toda chain perturbed with conservative noise was studied in \cite{StolzJSP10,CedricCMP2014} where again it was seen that the current decays with system size (anomalously).  
In \cite{BenettinJSP13} it was pointed out that  the Fermi-Pasta-Ulam (FPU) chain can be studied as a perturbation of Toda chain and that they exhibit similar behavior at short times.

Another motivation for our study is from the context of recent studies on  
thermalization in integrable quantum systems. 
It has been shown that integrable quantum systems prepared in special initial conditions, relax to a state that can be described by Generalized Gibbs Ensemble (GGE), i.e. thermal equilibrium state is described by a distribution $P=e^{-\sum_n \lambda_n I_n}/Z(\{\lambda_n\})$, where $I_n$ are the conserved quantities of the system, $\lambda_n$ are corresponding Lagrange multipliers and $Z$ is the appropriate partition function \cite{Rigol16}. On the other hand, typical states and also typical energy eigenstates are described by the usual Gibbs' ensemble (with only temperature specified) \cite{goldstein06,nandy16}. 
An interesting question then, is to see how integrability shows up in the dynamics of the system when it is prepared in an initial thermal Gibbs state.

In this paper we investigate the spatio-temporal equilibrium correlations of fluctuations of the three conserved quantities: stretch, momentum and energy  in 
the Toda chain. The equilibrium state is chosen to correspond to the one with specified temperature ($T$) and pressure ($P$) with zero average momentum. Our main results are as follows

(i) In all parameter regimes we find from numerical simulations that the correlations exhibit ballistic scaling, which means that all correlation functions have the form $C(x,t)= (1/t) f(x/t)$, where $f$ is some scaling function (non-universal).  

(ii) In two limiting cases the Toda system reduces   to the harmonic chain
and the hard particle gas. In these cases we are able to compute all correlation functions exactly. We show that there is excellent agreement between  direct simulations of the Toda with these exact results.

(iii) We follow the prescription used in the theory of fluctuating hydrodynamics of non-integrable anharmonic chains and carry out a transformation to the three ``normal'' modes corresponding to the three conserved quantities. 
We find that one can then again see a separation of the heat and sound modes, but unlike the non-integrable case, here the cross correlations between different modes are non-vanishing even at long times.


The plan of the paper is as follows. In Sec.~(\ref{model}) we precisely define the Toda chain model and gives a summary of some known  exact results. 
The numerical results for spatio-temporal correlations of the three conserved quantities in the  Toda chain  are presented in Sec.~(\ref{Eqb}). 
We also discuss the form of correlation functions in the normal-mode basis. 
We summarize the main findings of this work in Sec.~(\ref{conc}).

\section{Toda chain: Model, definitions  and summary of some exact results}
\label{model}
We first define the Toda model on a ring geometry. We consider $N$ particles with position $q_x$, momentum $p_x$ with $x=1,\ldots,N$.  We define a ``stretch'' variable $r_x = q_{x+1}-q_x$.  
The Toda Hamiltonian is given by
\begin{align}
&H = \sum_{x=1}^N \frac{p_x^2}{2} + V( r_x)~, \\ 
&{\rm where}~~V(r_x)=\frac{a}{b}e^{-b r_x}~, \nonumber
\end{align}
and we take periodic boundary conditions $q_{N+1}=\sum_{x=1}^N r_x=q_1+L$, $q_0=q_N-L$, where $L$ is the length of the lattice. 
The equations of motion are
\begin{equation}
m \ddot{q}_x= - a [~ e^{-b (q_x-q_{x-1})}-e^{-b (q_{x+1}-q_{x})}~ ]~, ~~~x=1,\ldots,N~. \label{EOM}
\end{equation}
Using the Lax pair formulation one can construct $N$ constants of motion, the first few of which are 
\begin{align}
I_1 &= \sum_{x=1}^N p_x~,  \label{IToda}
I_2 =  \sum_{x=1}^N \left[ \frac{p_x^2}{2} + V(r_x)\right]\\
I_3 &=  \sum_{x=1}^N \left[ \frac{p_x^3}{3} + (p_x+p_{x+1}) ~V(r_{x}) \right]\nonumber.
\end{align}
In addition  we have a trivial but important conserved quantity  $I_0 \equiv L = 
\sum_{x=1}^N r_x$, in the case of periodic boundaries.

{\bf{Limiting cases}}: If one takes the limit $b \to 0$, $a \to \infty$ with $a b = \omega^2$ constant, then one gets a harmonic chain with spring constant $\omega^2$. In addition there is a large linear term which can be canceled with an appropriate ``pressure'' term [adding a term $P r$ to the potential $V(r)$ with $P = a$].  On the other hand in the limit $b \to \infty$ the potential vanishes for $r >0$ and is infinite at $r=0$, thus mimicking a hard-particle gas. As we will see, in these limiting cases, all dynamical correlations can be exactly computed. In both these cases, some  equilibrium dynamical results  were already known 
\cite{JepsenJMP65, Montroll60, DharSabhapanditJSP15,HegedeSabhapanditPRL13}  and  even many exact properties of the non-equilibrium steady state are known \cite{RiederLebowitzJMP67,RoyDhar08,DharAdvPhy08}.
 
{\bf{Solitons and phonons}}: As noted in \cite{ShastryYoungPRB10} the Toda chain on the ring has a family of  the so-called ``Cnoidal" wave solutions that are periodic in time and space, very similar to the normal modes of a harmonic chain. For harmonic lattice the overall amplitude of the normal modes is a free parameter and apart from this freedom, there are exactly $N$ independent periodic solutions each specified by a wave-vector $k$ and a corresponding frequency $\omega_k$ (independent of amplitudes). For the nonlinear Toda lattice, one can again construct $N$ solutions specified by wave-vectors $k$ but there is a free ``non-linearity'' parameter depending on the  amplitude $A$ of the solution and in this case, the frequencies depend on $A$.  The explicit solutions are stated in \cite{ShastryYoungPRB10}. Here we note the observation made there, that for small amplitudes, the Cnoidal waves look like sinusoidal waves or phonons (the normal modes of a harmonic lattice) while for large amplitudes, they look like trains of solitons (localized excitations).

In the hard particle gas limit, the dynamics consist of particles moving ballistically and exchanging velocities on collision. A velocity pulse would simply pass un-scattered through this system. Thus this limit is characterized by ``non-interacting" solitons. So we see 
that the two limiting cases discussed above correspond to excitations being either phonon-like or soliton-like and for general parameters, we expect a mixture 
of these two.

{\bf Specification of the equilibrium state and definition of correlation functions}:
The Toda chain has a large number of conserved quantities, and accordingly one can construct generalized ensembles which are invariant 
distributions. Such general ensembles are specified by a set of $N$ Lagrange multipliers corresponding to the $N$ conserved quantities. Here we restrict our discussion to the special case where the initial state is 
prepared such that only the conserved quantities energy, stretch and momentum are specified while all other Lagrange multipliers are set to zero. 
More specifically we prepare the system initially in a state described by the following canonical ensemble (with zero average momentum) and  at specified temperature $T$ and pressure $P$:
\begin{equation}
Prob(\{ r_x,p_x\})=  \frac{ e^{-\beta  \sum_{x=1}^N \left[ p_x^2/2+V(r_x)+P r_x\right]}}{Z}~,\label{Peq}
\end{equation}
where the partition function is simply given by $Z=[\int_{-\infty}^\infty dp \int_{-\infty}^\infty dr e^{-\beta (p^2/2+V(r)+Pr)}]^N $. 

Corresponding to the three global conserved quantities 
$(I_0,I_1,I_2)$, we can define the local conserved fields $r_x(t),p_x(t),e_x(t)=p_x^2/2+V(r_x)$. It is easy to see that they  satisfy the continuity equations
\begin{align}
\partial_t r_x &=p_{x+1}-p_x \nonumber\\
\partial_t p_x &= V'(r_x)-V'(r_{x-1}) \label{conteq}\\
\partial_t e_x &= p_{x+1}V'(r_x)-p_x V'(r_{x-1})~.\nonumber
\end{align}
Defining a local pressure variable ${P} = - V'(r)$, and the discrete derivative $\partial_x f(x)= f(x+1)-f(x)$ we see that the above equations can be written in the following form
\begin{align}
\partial_t r_x(t) +\partial_x j_r(x,t)& = 0~, \label{conteq2} \\
\partial_t p_x(t) +\partial_x j_p(x,t)& = 0~, \nn \\
\partial_t e_x(t) + \partial_x j_e(x,t) &= 0~,~~{\rm where}\nn \\
[j_r(x,t),j_p(x,t),j_e(x,t)]&=[-p_x(t), {P}_{x-1}(t), p_x(t) {P}_{x-1}(t)] \nn   
\end{align}
Next, we define the fluctuations of the fields from their equilibrium values as
\begin{equation}
u_1(x,t)=r_x(t)-\la r \ra,~u_2(x,t)=p_x(t),~u_3(x,t)=e_x(t)-\la e \ra~, \label{eqcorr1}
\end{equation}
where $\la \ldots \ra$ denote average over the initial equilibrium state.  
We will look at the following  dynamic correlation functions: 
\be
{C}_{\alpha \nu} (x,t) = \langle u_{\alpha}(x,t) u_{\nu}(0,0) \rangle~, \label{eqcorr2}
\ee
with $\alpha,\nu=1,2,3$. The average is over initial conditions chosen from Eq.~(\ref{Peq}) and the dynamics in Eq.~(\ref{EOM}) [or equivalently the first two equations in Eq.~(\ref{conteq})].

{\bf {Correlation functions in the special limiting cases of Toda lattice}}: 
Exact results for the  correlations of velocity $\la p_x(t) p_0(0) \ra$    were obtained in the papers by Montroll and Mazur \cite{Montroll60} for the harmonic chain  and by Jepsen \cite{JepsenJMP65}  for the hard particle gas.
The dynamics of harmonic crystal being linear and the initial conditions taken from Gaussian distribution makes it simple to obtain exactly the full set of  correlations $C_{\alpha\beta}(x,t)$. It turns out that for  the hard-particle gas, one can use a recently developed formalism \cite{DharSabhapanditJSP15}, to again compute the full set of correlation functions \cite{Kunduetal16}. Here we summarize these results (some details of the calculations are given in the appendix).  

For a harmonic chain with nearest neighbor spring constant $\omega^2$, the correlation functions are given by
\bea
C_{rr}(x,t) &=& T {\cal J}_{2|x|} (2\omega t) /\omega^2 \label{harmC} \\
C_{rp}(x,t) &=& T \left[-\frac{{\cal J}_{2|x|-1} (2\omega t)}{\omega}\theta(-x)  + \frac{{\cal J}_{2|x|+1} (2\omega t)}{\omega}\theta(x)\right] \nn \\
C_{pr}(x,t) &=&  T \left[ -\frac{{\cal J}_{2|x|+1} (2\omega t)}{\omega}\theta(-x)  + \frac{{\cal J}_{2|x|-1} (2\omega t)}{\omega}\theta(x)\right] \nn \\
C_{pp}(x,t)  &=&  T {\cal J}_{2|x|} (2\omega t) \nn\\
C_{ee}(x,t)  &=& \frac{1}{2}\left[ C^2_{rr}(x,t)  + C^2_{rp}(x,t)  + C^2_{pr}(x,t)  + C^2_{pp}(x,t)\right] \nn
\eea
where ${\cal J}_n(z)$ is Bessel function of the first kind of order n and $\theta(x)$ is the Heaviside theta function.

For the hard particle gas, we consider initial velocities chosen from Maxwell distribution with variance $\bar{v}^2 =T$.  The correlation functions are then given by
\bea
C_{rr}(x,t)  &=& \frac{1}{\rho^2 \sigma_t}  \frac{e^{-\frac{1}{2}({\frac{x}{\sigma_t}})^2}}{\sqrt{2 \pi}} \label{HPGC} \\
C_{pp}(x,t) &=& \frac{\bar{v}^2}{\sigma_t} \left({\frac{x}{\sigma_t}}\right)^2 \frac{e^{-\frac{1}{2}({\frac{x}{\sigma_t}})^2}}{\sqrt{2 \pi}}\nn \\
C_{ee}(x,t) &=& \frac{\bar{v}^4}{4\sigma_t} \left[ \left( \frac{x}{\sigma_t}\right)^4 - 2\left( \frac{x}{\sigma_t}\right)^2 + 1 \right]  \frac{e^{-\frac{1}{2}({\frac{x}{\sigma_t}})^2}}{\sqrt{2 \pi}}\nn
\eea
 where $\rho = P/T$ is the average density and $\sigma_t = \rho \bar{v} t$.

{\bf Sum rules}: We note here \cite{SpohnJStat14,Book} that  the correlation functions of conserved quantities satisfy the following exact sum rules (see appendix for derivation), in the limit $N \to \infty$
\eqa{
\sum_x C^{\alpha\beta}(x,t) &=  \sum_x C^{\alpha\beta}(x,0)~, \label{sumrules} \\
 \frac{d}{d t} \sum_x  x C^{\alpha\beta}(x,t) &= \la J^\alpha(0) u^\beta(0,0) \ra~, \nn \\
\frac{d^2}{d t^2} \sum_x  x^2C^{\alpha\beta}(x,t) &= 2 \la J^\alpha (t) j^\beta (0,0)\ra \\
{\rm where}~~J(t)& =\sum_x j(x,t)~. \nn
}	
These sum rules serve as useful check of numeric simulations. Further they provide useful information on transport properties. For example, the last of the above equation enables one to relate total current correlations to spreading of correlation functions of corresponding conserved quantities. One can then try to say
something about non-equilibrium transport via linear response theory \cite{liu,li}.  For the case of the integrable models studied here, we see ballistic scaling of correlations of all conserved currents, and this immediately implies that 
the corresponding total currents do not decay to zero in the infinite time limit.   

We now first present results from direct simulations on the form of these different correlation functions. 
In our simulations we explore different parameter regimes and in the two limiting cases, compare our results with the above exact results.

\section{Numerical results for correlations of conserved quantities}
\label{Eqb}

\begin{figure}
\includegraphics[width=0.5\textwidth,height=\textheight,keepaspectratio ]{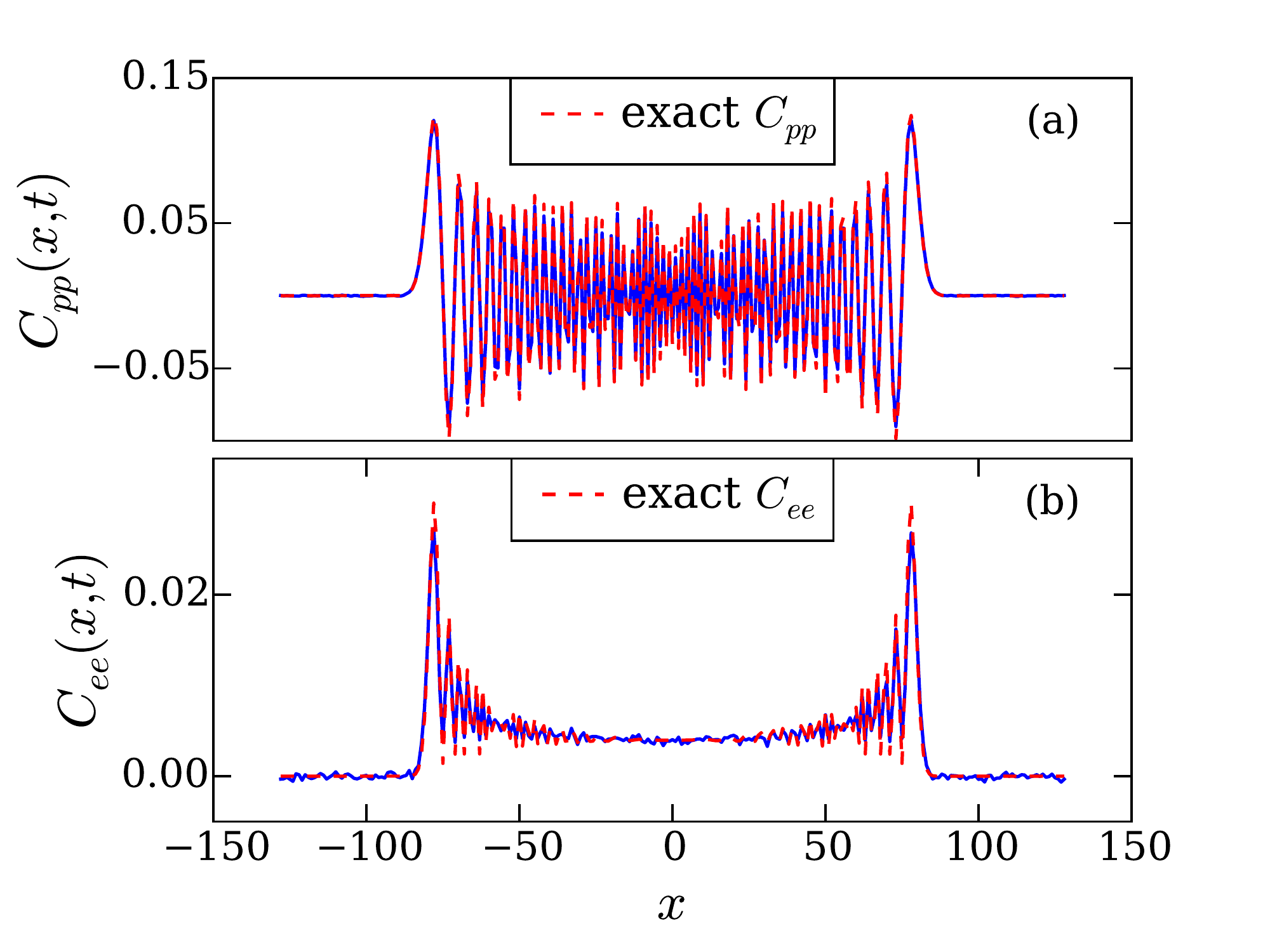}
\includegraphics[width=0.5\textwidth,height=\textheight,keepaspectratio ]{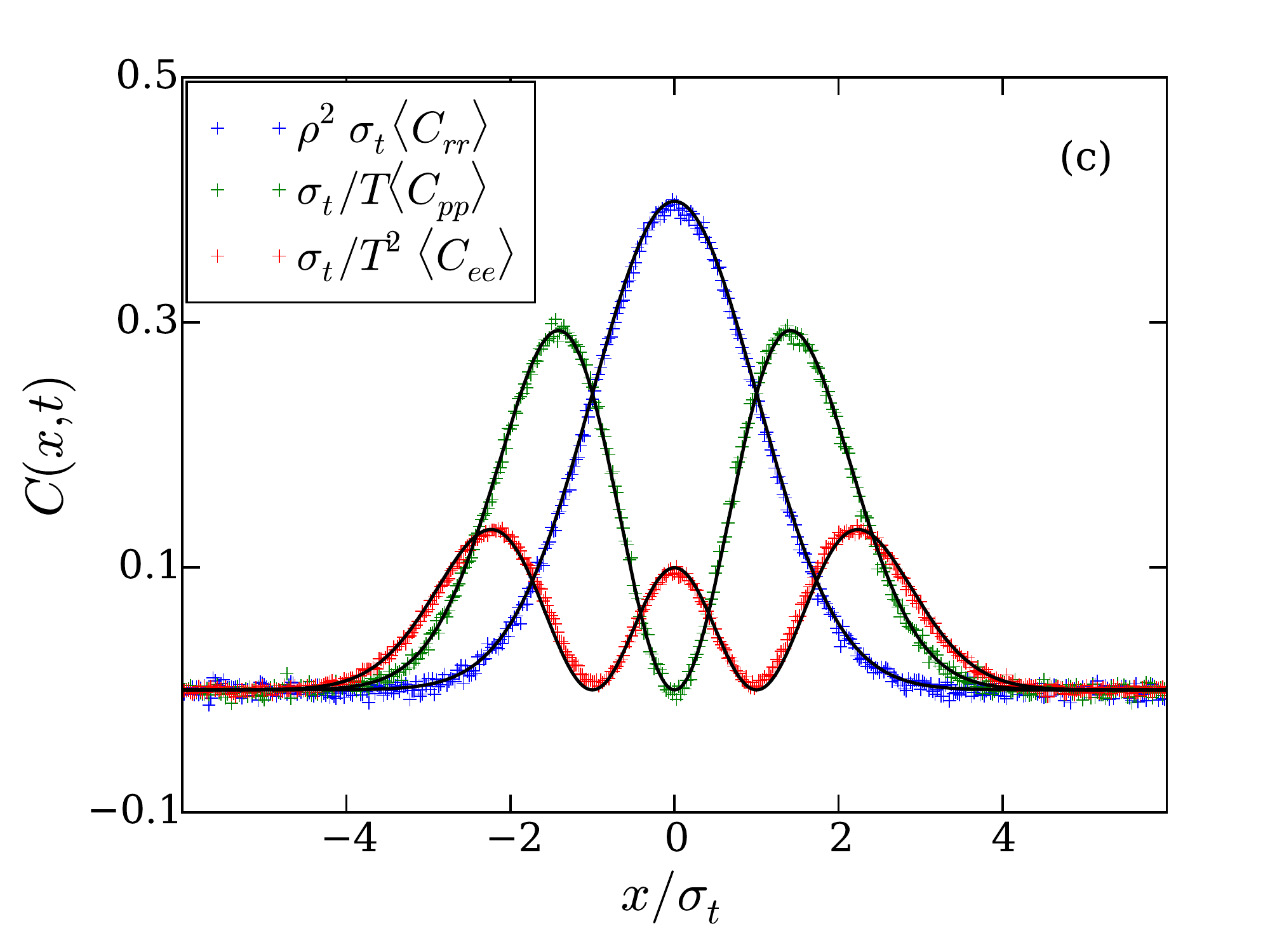}
\caption{(a,b) Toda chain with parameters $a = 20.0, b = 0.05,  P=20.0, T=1.0, N=256$ at time $t=80$. This corresponds to the harmonic limit. The simulations of the Toda are compared  with the exact harmonic chain correlation functions (red dashed lines) as given in text. Here $\omega^2=1$, hence $C_{rr} = C_{pp}$.  (c) Toda chain with parameters $a = 0.1, b = 10.0, P = 0.1, T = 1.0, N = 1024$ corresponding to hard-particle limit, at time $t = 400$. The solid black lines are the values of exact correlation function , as given in text (with $\sigma_t = \rho \bar{v}t$).}
\label{exactcorr}
\end{figure}

{\bf Numerical details:} The Toda-chain is simulated by numerically evaluating Eq.~(\ref{EOM}) using the velocity-Verlet algorithm. We choose a small time-step ($dt \leq 0.01$) in the simulations which keeps the total energy and momentum constant to a high accuracy ( relative error less than  $10^{-6} $ for energy and $10^{-4} $  in $I_3$ ).
To capture the equilibrium correlations, we prepare the system in an initial state in a canonical $(T,P)$ ensemble by drawing random numbers ${p,r}$ for each particle from the distribution  ${e^{-\beta (p^2/2 +V(r)+P r)}}/{Z}$ through inverse transform sampling. For the partition function to be bounded, we require that pressure is non-zero for Toda Lattice. The  full set of spatio-temporal correlation functions, defined in Eq.~(\ref{eqcorr2}) are computed by taking averages over $10^6 - 10^7$ initial conditions.

\begin{figure}
\includegraphics[width=0.5\textwidth,height=\textheight,keepaspectratio ]{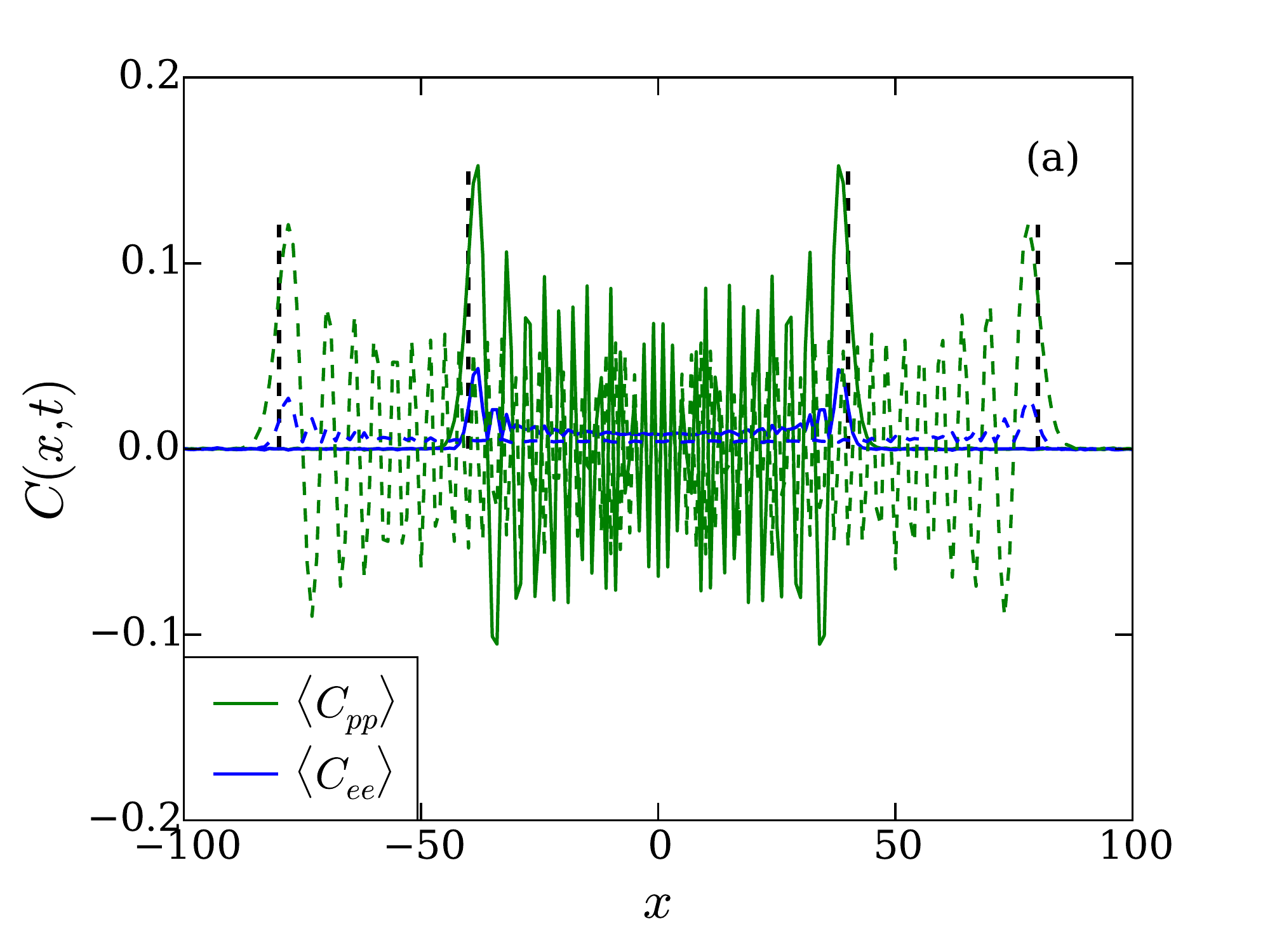} 
\includegraphics[width=0.5\textwidth,height=\textheight,keepaspectratio ]{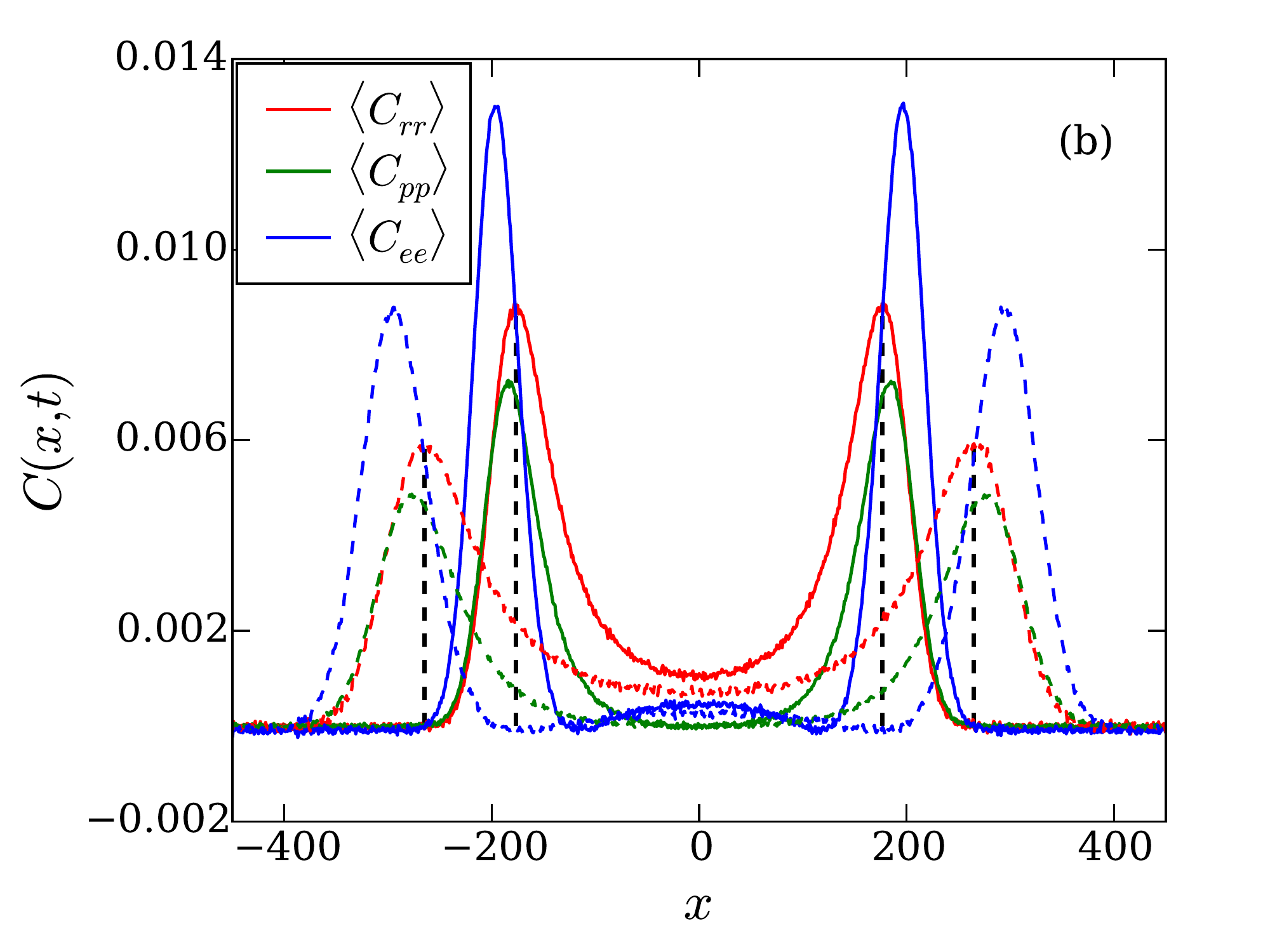} 
\includegraphics[width=0.5\textwidth,height=\textheight,keepaspectratio ]{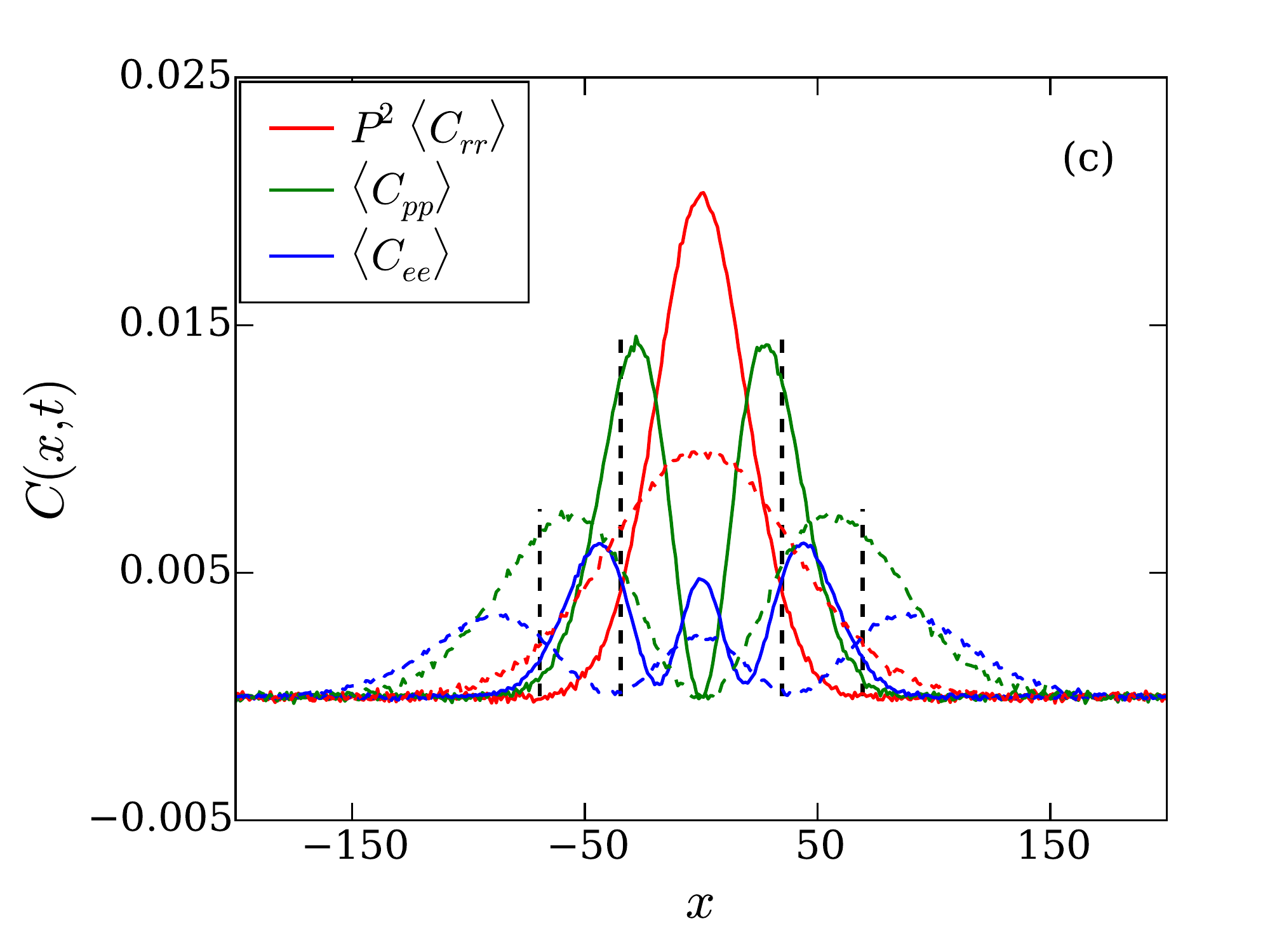} 
\caption{Diagonal correlation functions Eq.~(\ref{eqcorr2}) for Toda lattice for two different times (dashed line indicates later time). The black dashed line show position of sound velocity as predicted from Eq.~(\ref{soundspeed}) at the two times (a) Harmonic Limit at time $t = 80, 120$ [parameters of Fig.~\ref{exactcorr}(a)].    (b) Toda chain with parameter $a = 1, b = 1, P = 1, T = 1 ,N = 1024$ at time $t = 200,300$. (c) Hard-Particle limit  [parameters of Fig.~\ref{exactcorr}(c)] at time $t = 200, 400$.  }
\label{directcorr}
\end{figure}

We present numerical results and discuss their scaling  in  three different parameter regimes. These correspond to the harmonic and hard particle limits and an intermediate regime.  In the former cases, comparisons are made  with the exact results stated in the previous section. For the three conserved quantities we will use the notation $r \leftrightarrow 1,~p \leftrightarrow 2,~e \leftrightarrow 3$.

\begin{figure}
\includegraphics[width=0.55\textwidth,height=\textheight,keepaspectratio ]{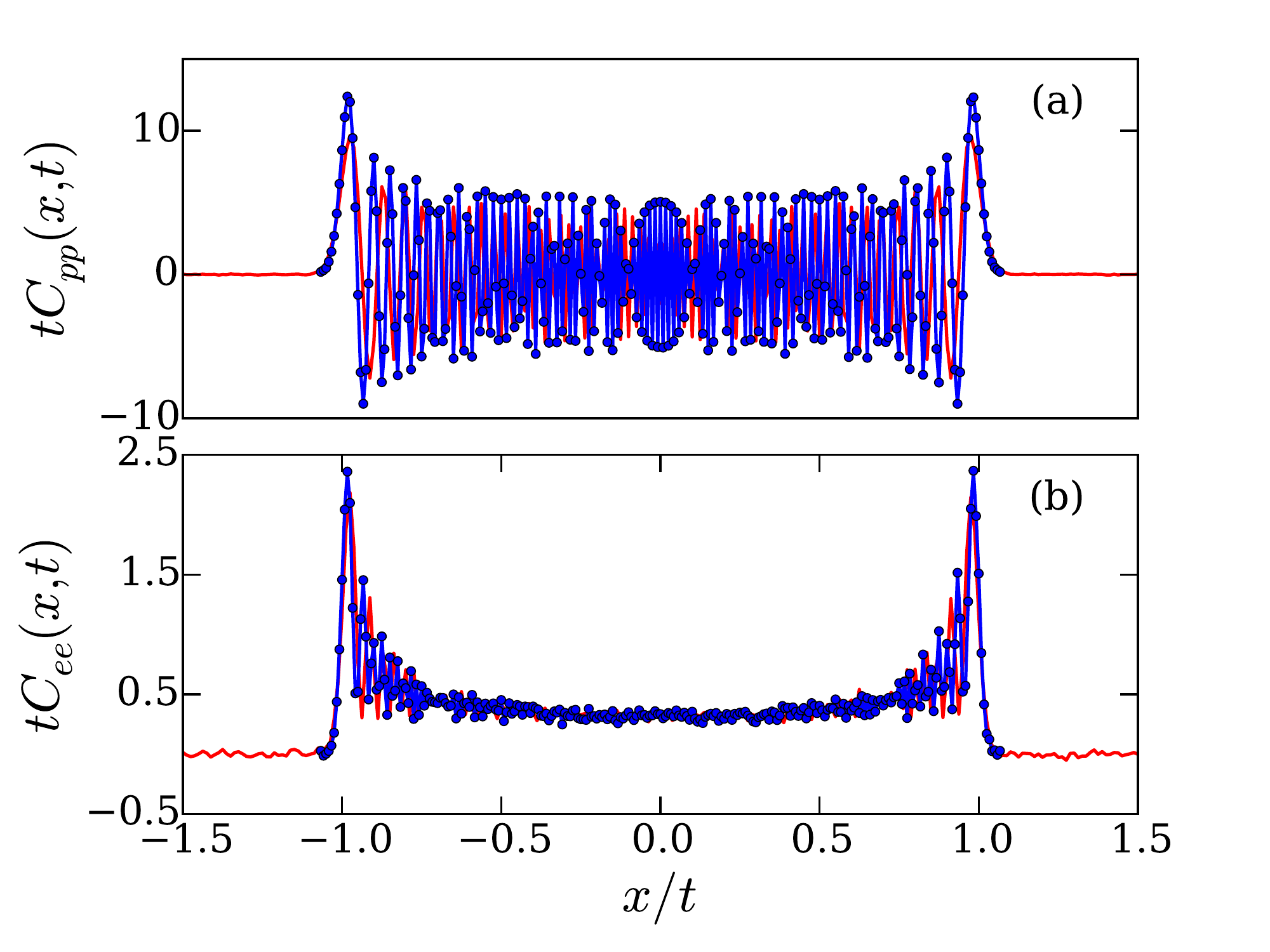}
\includegraphics[width=0.55\textwidth,height=\textheight,keepaspectratio ]{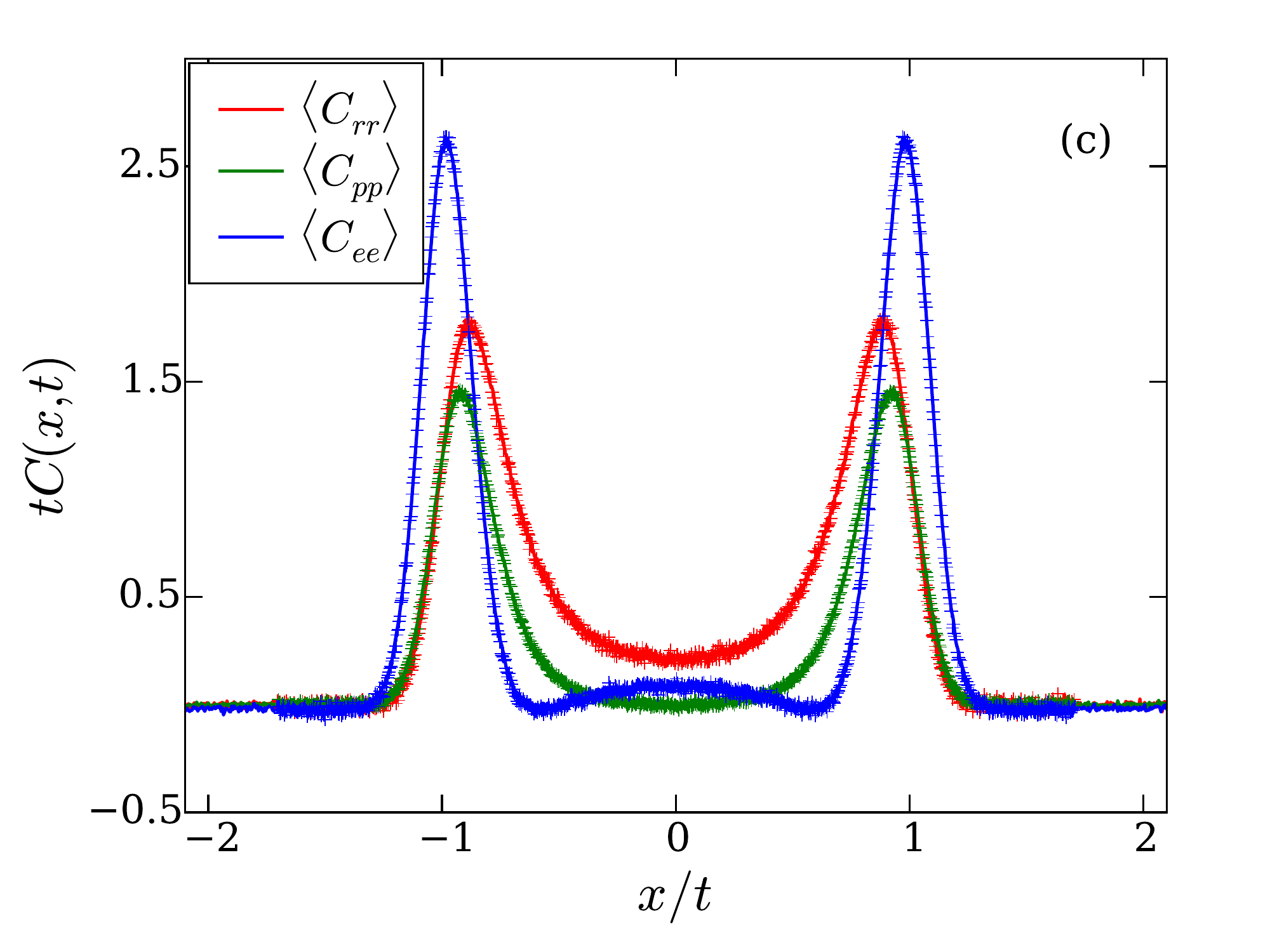} 
\includegraphics[width=0.55\textwidth,height=\textheight,keepaspectratio ]{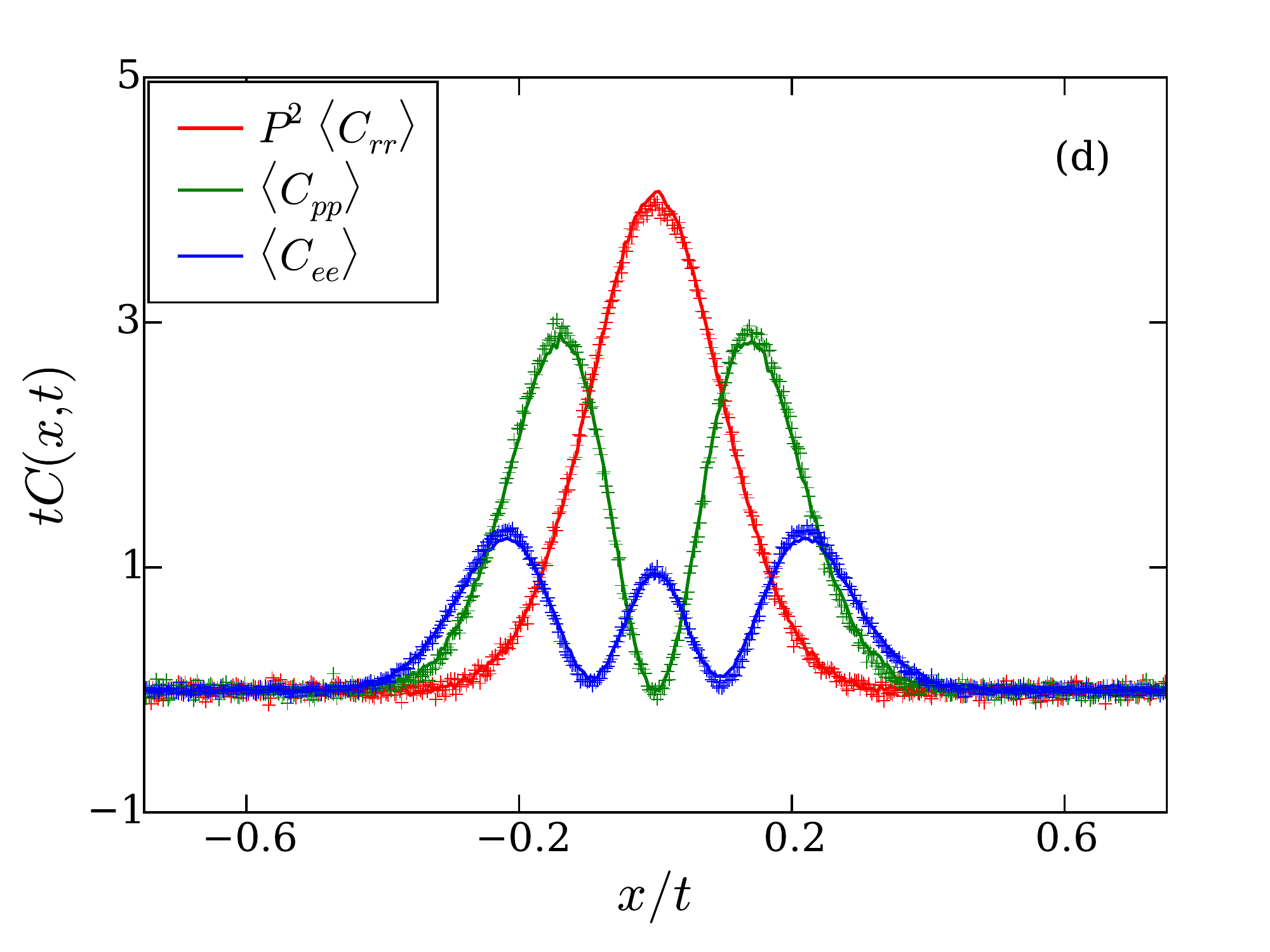} 
\caption{The diagonal correlation functions Eq.~(\ref{eqcorr2}) for Toda lattice in various limits for two different times are plotted with ballistic scaling. (a) and (b) shows collapse momentum and energy correlations respectively in harmonic limit with parameters that of Fig.~\ref{directcorr}(a). Although the scaling is good in the bulk, the edges show significant deviations. (c) shows ballistic scaling for Toda with parameters as that of Fig.~\ref{directcorr}(b). In (d) we show ballistic scaling in the hard particle limit with parameters as in Fig.~\ref{directcorr}(c).}
\label{directcorrscaled}
\end{figure}

\textbf{Case I: $a=20, b=0.05,  P=20.0, T=1.0$} ---
In this limit, the Toda lattice is expected to show similar characteristics of the harmonic lattice. In  Fig.~\ref{exactcorr}(a) and \ref{exactcorr}(b) we show results for the diagonal correlations $C_{rr},C_{ee}$ in the Toda lattice respectively and compare them with  exact  harmonic chain results as given in Eqs.~(\ref{harmC}). We find an excellent agreement. For our parameters, the effective spring constant $\omega^2=1$ and hence $C_{rr}=C_{pp}$. 
The correlations are extended and  oscillatory. In Fig.~\ref{directcorr}(a), the momentum and energy spatio-temporal correlations are shown for two different times illustrating how they spread with time. The speed of sound here is $c\approx 1$. In Fig.~\ref{directcorrscaled}(a) and \ref{directcorrscaled}(b), we plot the same data after scaling the $x$ and $y$ axes by factors of $1/t$ and $t$ respectively (ballistic scaling).
We see a good collapse in the bulk with some deviations near the sound peaks,
which occur near the edge.

\textbf{Case II:  $a=1.0, b=1.0, P=1.0, T=1.0$} ---
This corresponds to the intermediate regime and we no longer see  the oscillations in the correlations.  In Fig.~\ref{directcorr}(b), the momentum and energy  correlations are shown for two different times. The speed of sound here is $c = 0.8833...$.  The stretch and momentum correlations only have peaks at the edges,
while  the energy correlation has an additional  small peak in the middle.
In Fig.~\ref{directcorrscaled}(b) we see that there is  a very  good ballistic scaling of the correlation functions.

\textbf{Case III: $a=0.1, b=10.0,  P=0.1, T=1.0$} ---
This corresponds to the hard particle gas limit and again we see no oscillations. In Fig.~\ref{exactcorr}(c) the results for correlation functions from direct simulations of the Toda chain are compared with the exact results for the hard 
particle gas in Eq.~(\ref{HPGC}). We again see excellent agreement with the numerical data. We now see that the nature of correlations are very different. The stretch correlation has a single peak at the center and energy 
correlation has a relatively large central peak. 
In Fig.~\ref{directcorr}(c) the  correlations are shown at two different times.
The speed of sound is $c=0.1709...$ while Fig.~\ref{directcorrscaled}(c) shows the expected ballistic scaling.

\begin{figure}
\includegraphics[width=0.5\textwidth,height=\textheight,keepaspectratio ]{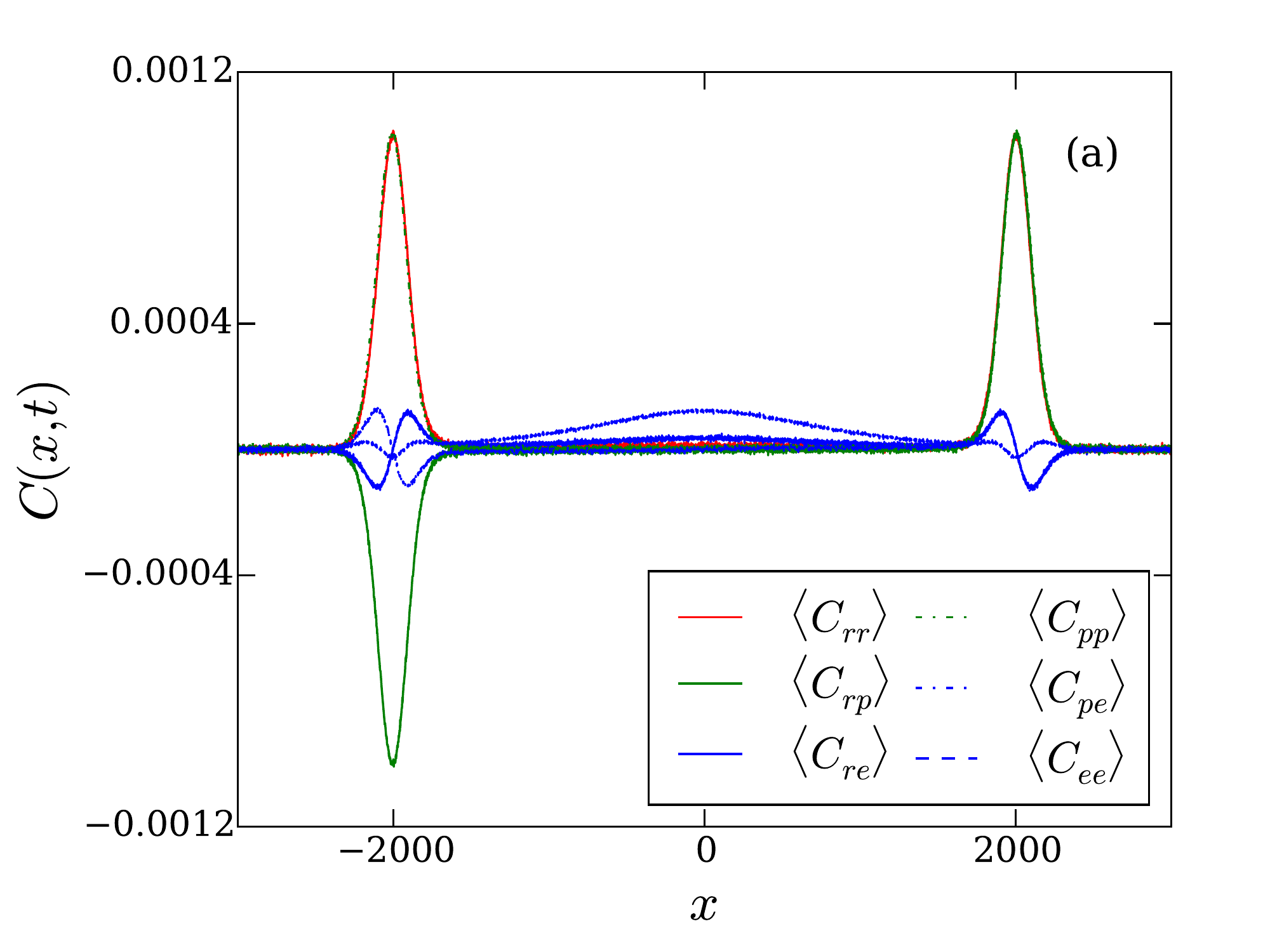}
\includegraphics[width=0.5\textwidth,height=\textheight,keepaspectratio ]{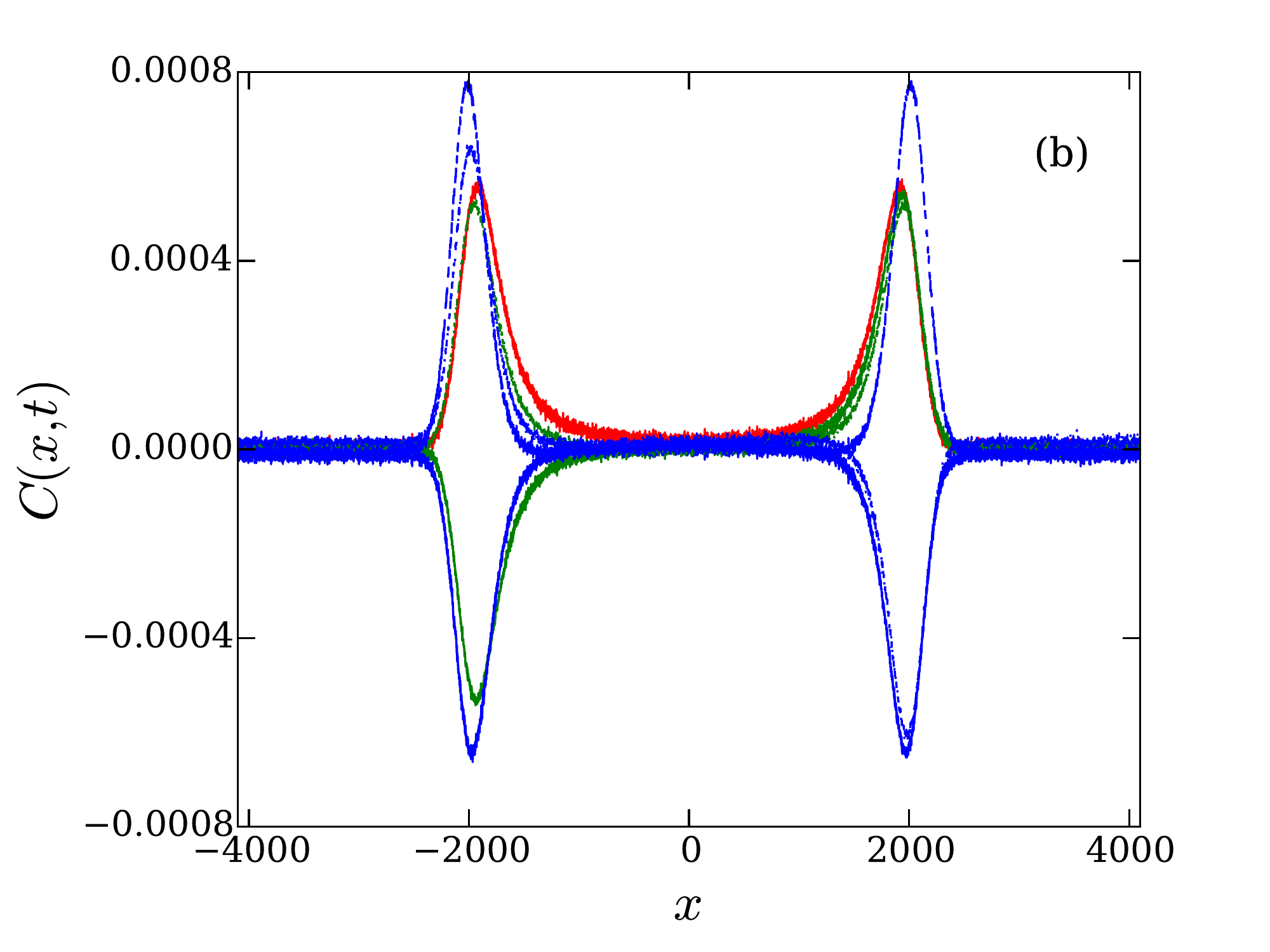}
\caption{Figure shows all correlation functions between the three locally conserved quantities $(r,p,e)$ for (a) truncated Toda chain [potential given by Eq.~(\ref{TruncTodaPot})] and parameters $P = 0$, $T = 0.5$, $N = 8192$ at time $t = 2000$. (b) Toda chain with $a=1, b=1, P=1$ and all other parameters the same as in (a).}
\label{AllCorr}
\end{figure}

To see the huge effect of integrability on the form of correlations, we show 
results from a simulation with a potential corresponding to a truncated Toda potential with parameters $P,T$ chosen to be close to the actual Toda simulations. 
Note that truncation leads to an FPU potential and is expected to destroy integrability. 
The truncated Toda potential to quartic order is given by
\be V_{tr}(r) = \frac{r^2}{2} - \frac{r^3}{6} +\frac{r^4}{24}~,\label{TruncTodaPot}\ee
and we have set $P=0.0, T=0.5$ to match the equilibrium properties with that of Toda chain with parameters $a=1, b=1, P=1, T=0.5$. With this parameters, the speed of sound in truncated Toda chain is $c = 1.004...$ and for Toda chain is $c = 0.938...$ which is about a $6 \%$  difference.

In  Fig.~\ref{AllCorr} we show a comparison of the correlation functions of Toda chain with the corresponding truncated Toda chain.  We see that they
show significant qualitative  differences.   
In particular for the truncated Toda (FPU) chain, the correlation functions show localized and  well-separated peaks, while in the Toda chain, they are broad and overlapping. The cross correlations are of similar order in both cases and we will now see how this changes when we transform to normal mode basis. The normal mode representation more clearly shows the difference between the Toda results and the FPU.


{\bf Description in terms of  normal modes}:
In the usual hydrodynamic theory of anharmonic chains \cite{SpohnJStat14}, 
it is convenient to go to a description in terms of ``normal'' hydrodynamic modes of the system. The normal modes, which we will denote by $(\phi_+,\phi_0,\phi_-)$ consist of linear combinations of the original field $(u_1,u_2,u_3)$ chosen in such a way that the correlation matrix becomes approximately diagonal at long times, i.e, the cross correlations between different modes become negligible at long times. At the level of linearized hydrodynamics, for the diagonal elements of the correlation 
matrix, well-separated peaks for each mode is seen. Specifically 
one finds (at the linear level) a single diffusively spreading heat mode and two  propagating sound modes  moving with speeds $\pm c$. While it is not obvious
what such a normal mode transformation will achieve for our integrable system,
we nevertheless proceed to construct such a transformation (using the three 
variable description) and analyze the correlations in this basis.

We briefly review the construction of the normal mode transformation, starting 
with the microscopic continuity  equations given by Eq.~(\ref{conteq2}). 
The conserved currents $j_\alpha$ are then expanded about their equilibrium value up to linear order in the fields leading to the linear equations
\be
\partial_t u_\alpha(x,t)+\partial_x (A^{\alpha \beta}u_\beta(x,t)) =0 ,
\ee
where
\[
A=\left(
\begin{array}{ccc}
 0 & -1 &0 \\
 \partial_l P&0&\partial_e P \\
0 & P & 0
\end{array}
\right).
\]
The partial derivatives above are computed using the equilibrium equation of state $P=P(l,e)$ where $l=\la r\ra, e=\la e \ra$.
The diagonalization of the matrix $A$  leads to the form 
$RAR^{-1}=diag(-c,0,c)$, where the matrix $R$ is completely fixed by the normalization condition $RC(t=0)R^T=1$, with $C$ the correlation matrix. We refer the reader to \cite{SpohnJStat14} for explicit expressions. The constant $c$ corresponds to the sound velocity and can be computed explicitly from equilibrium correlation functions through the formula \cite{SpohnJStat14}
\bea
c^2 = \frac{1}{\Gamma} \left(\frac{1}{2\beta^2} + \la V + Py; V +Py \ra \right),
\label{sound}
\eea
with $\Gamma = \beta(\la y;y\ra\la V;V \ra- \la y;V\ra^2) + \frac{\la y;y \ra}{2\beta}\nn$, and where $\la A;B\ra=\la A B\ra-\la A\ra \la B\ra$. 
For the Toda potential one can simplify Eq.(\ref{sound}) to get the form
\be
c^2=\frac{b^2}{\beta}\frac{ \left(2 z^2 \psi ^{(1)}(z)-2 z+1\right)}{ ((2 z+1) \psi ^{(1)}(z)-2)},
\label{soundspeed}
\ee
where $z=\frac{\beta P}{b}$ and $\psi^{(1)}(z)$ is Polygamma function which is defined as $\psi^{(1)}(z) = \frac{d^2}{dz^2} \log(\Gamma(z))$ and $\Gamma(z)$ is the standard Gamma-function. It is interesting to note that for the special case with $P=b$, the above formula 
is very close to one derived in \cite{BauerMertensPhyB88}.

For small $b$  and $P=a$, Eq.~(\ref{soundspeed})  can be expanded to give the expected speed of sound in a harmonic chain  $c=\sqrt{ab}$.  
In the other limit, when $b \rightarrow \infty$ and the external pressure is $P$ the above formula  gives the hard particle limit  $c = \sqrt{3 \beta}P$. 
These two limits can also be obtained in the high temperature (corresponding 
to large $b$) and low temperature limits (small $b$) by expanding with respect 
to $z$, leading to the same expressions for speed of sound to the leading order.

The normal mode transformation is then defined by 
$\phi_s = \sum_\alpha R_{s,\alpha} u_\alpha$, for $s=+,0,-$. 
We can then compute correlations for these normal modes 
\be
C_{rs}= \la \phi_r(x,t) \phi_s(0,0)\ra~, 
\ee  
for $r,s=+,0,-$. As we will see this normal mode transformation separates the two sound modes $s=+,-$ moving with velocity $\pm c$ respectively and the heat mode $s=0$.  All the modes continue to show  ballistic scaling. 
We now show numerical data of the correlations in normal modes for the Toda chain in various parameter regimes.

\begin{figure}
\includegraphics[width=0.5\textwidth,height=\textheight,keepaspectratio ]{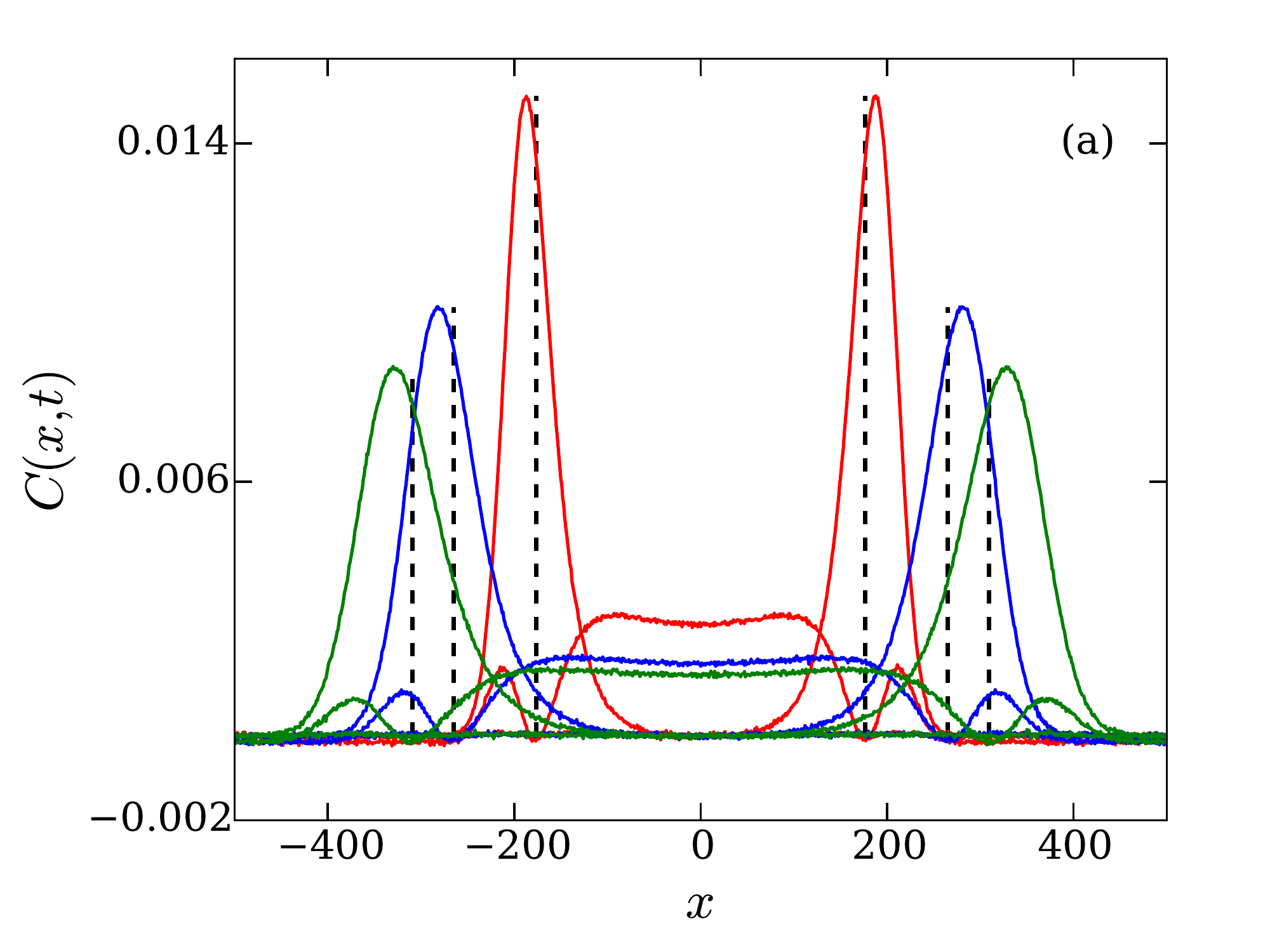}
\includegraphics[width=0.5\textwidth,height=\textheight,keepaspectratio ]{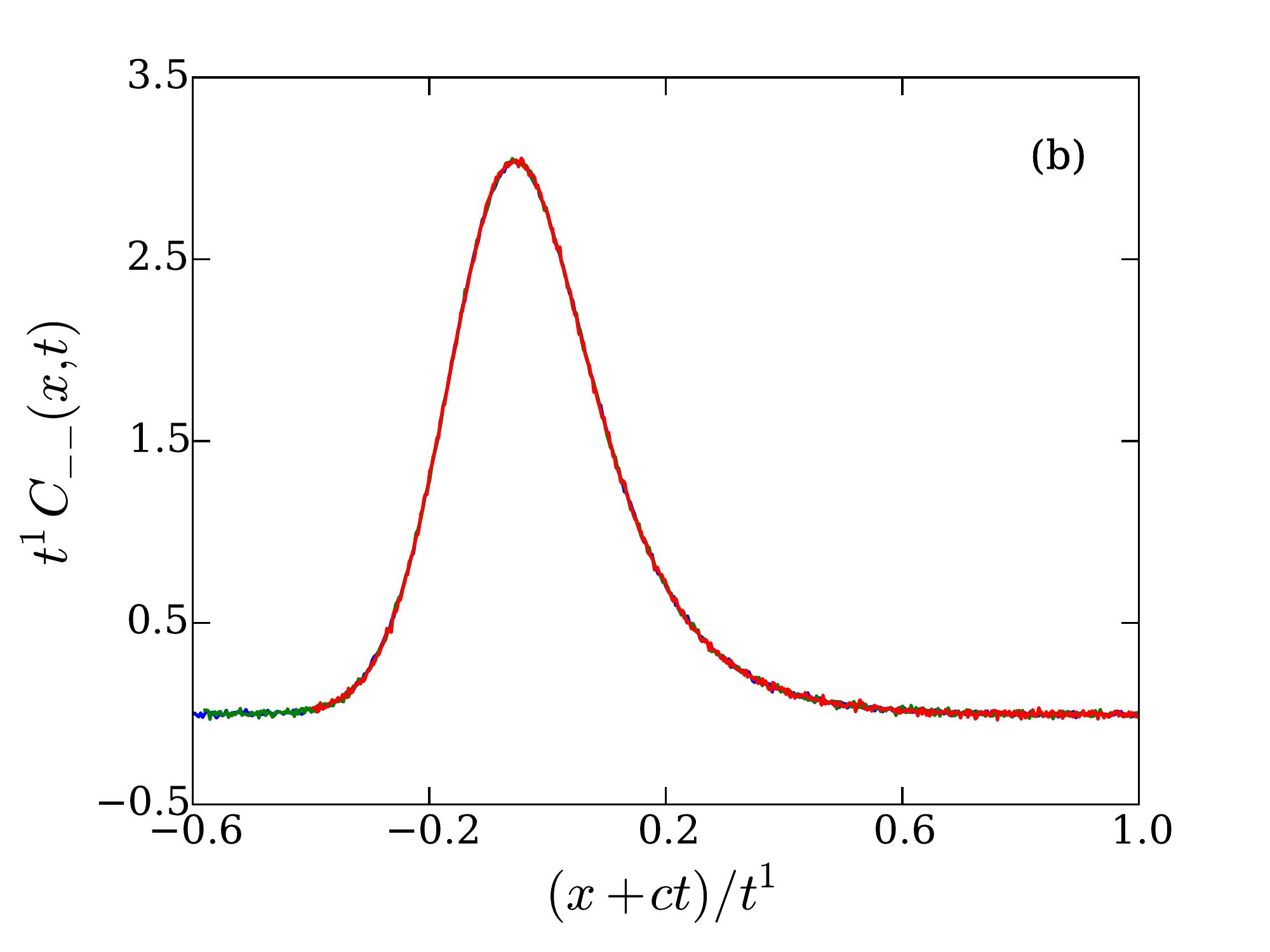}
\includegraphics[width=0.5\textwidth,height=\textheight,keepaspectratio ]{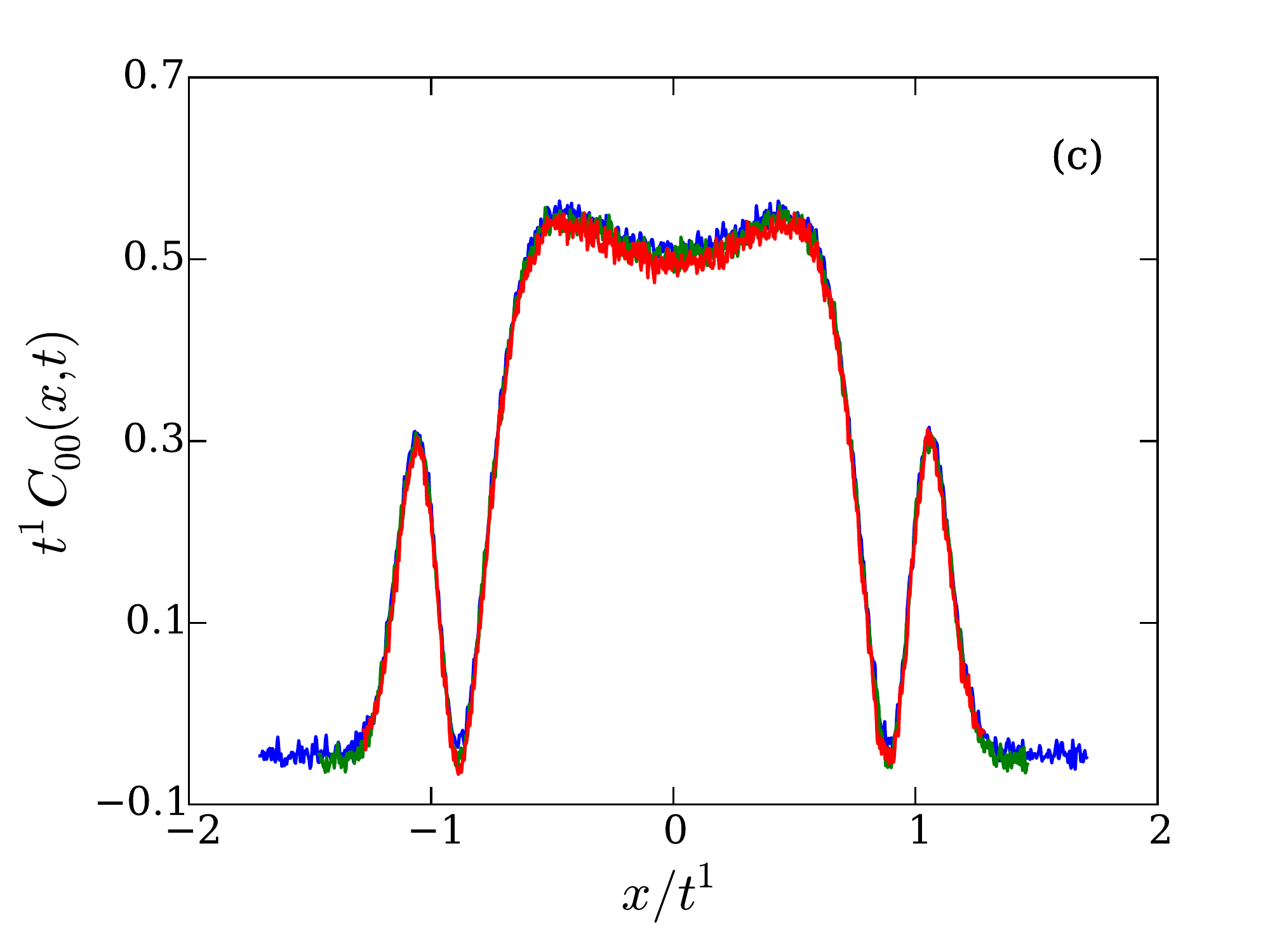}
\caption{Normal mode representation- Case I with parameters $a = 1, b = 1, P=1,  T = 1$ and $N=1024$.  (a) Heat and sound modes plotted together at times $t=200$ (red), $t=300$ (blue) and $t=350$ (green). The two sound modes  move to left and to the right with velocities $\mp c$ (vertical dashed lines indicate the distance $ct$). (b) The sound modes at the three times are scaled ballistically 
and we see a good collapse even at small times. (c) The heat modes  are scaled ballistically.}
\label{T1}
\end{figure}

{\textbf {Case I:} $a=1.0, b=1.0, P=1.0, T=1.0$} ---
In Fig.~\ref{T1}(a) we show the sound and heat modes plotted together at three different times $t=200, 300, 350$. The speed of sound is $c = 0.883...$.  The scaled right moving sound  modes and the  scaled heat modes are plotted in Fig.~\ref{T1}(b) and Fig.~\ref{T1}(c) respectively. 
The scaling collapse is very good even for short times. The sound mode is 
broad and asymmetric. The heat mode on the other hand has a broad central peak and also significant side peaks. The amplitude of heat mode is much less than that of sound mode, which implies less  scattering. In Figs.~\ref{T1}(b) and \ref{T1}(c) we show the ballistic scaling of the right moving sound mode and the heat mode. Note that the shift by $ct$ for the sound mode is not really necessary to see scaling collapse for the ballistic case. Typically we find that the off-diagonal correlations are of same magnitude as that of the diagonal correlations.

\begin{figure}
\includegraphics[width=0.5\textwidth,height=\textheight,keepaspectratio ]{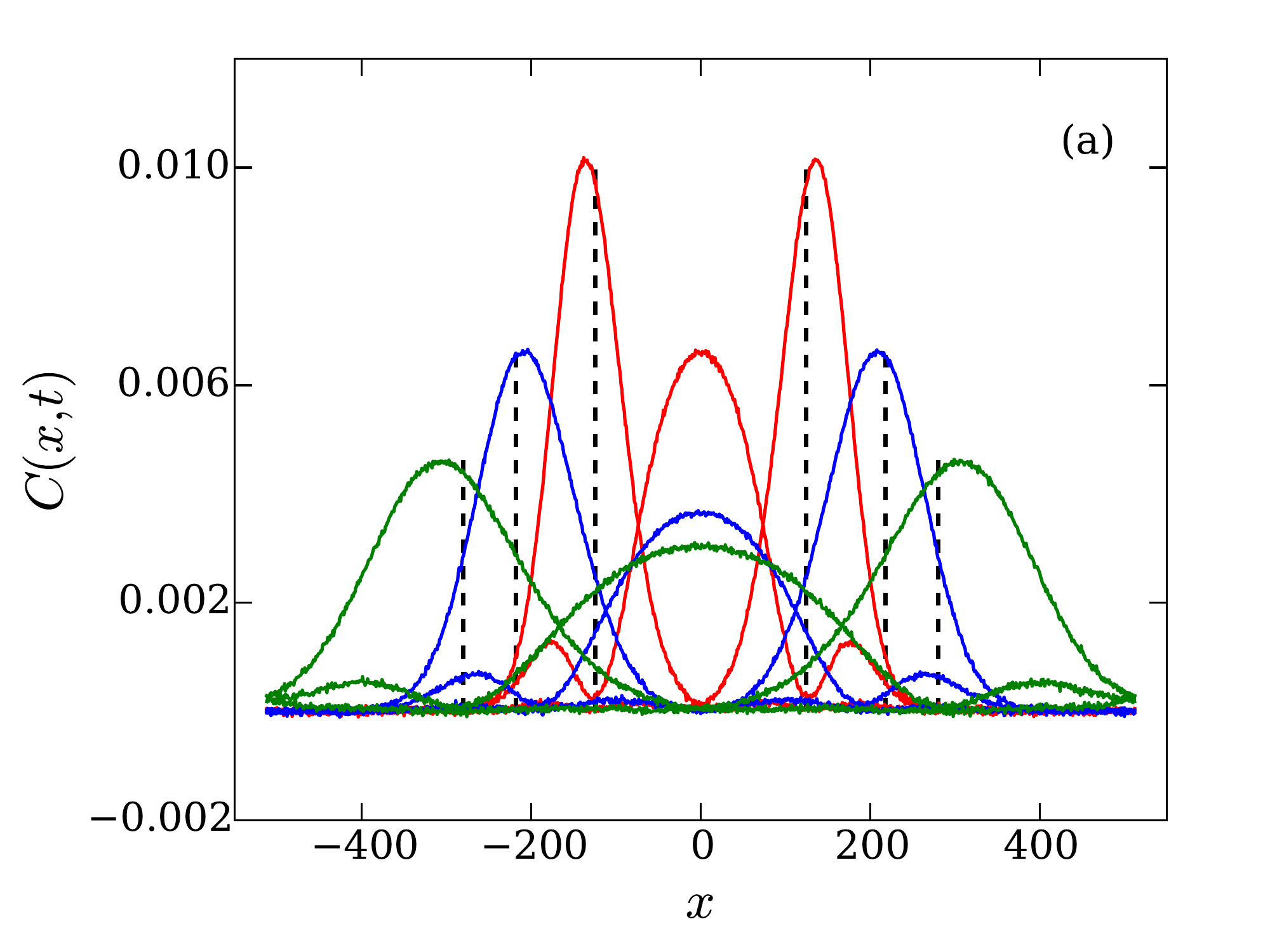}
\includegraphics[width=0.5\textwidth,height=\textheight,keepaspectratio ]{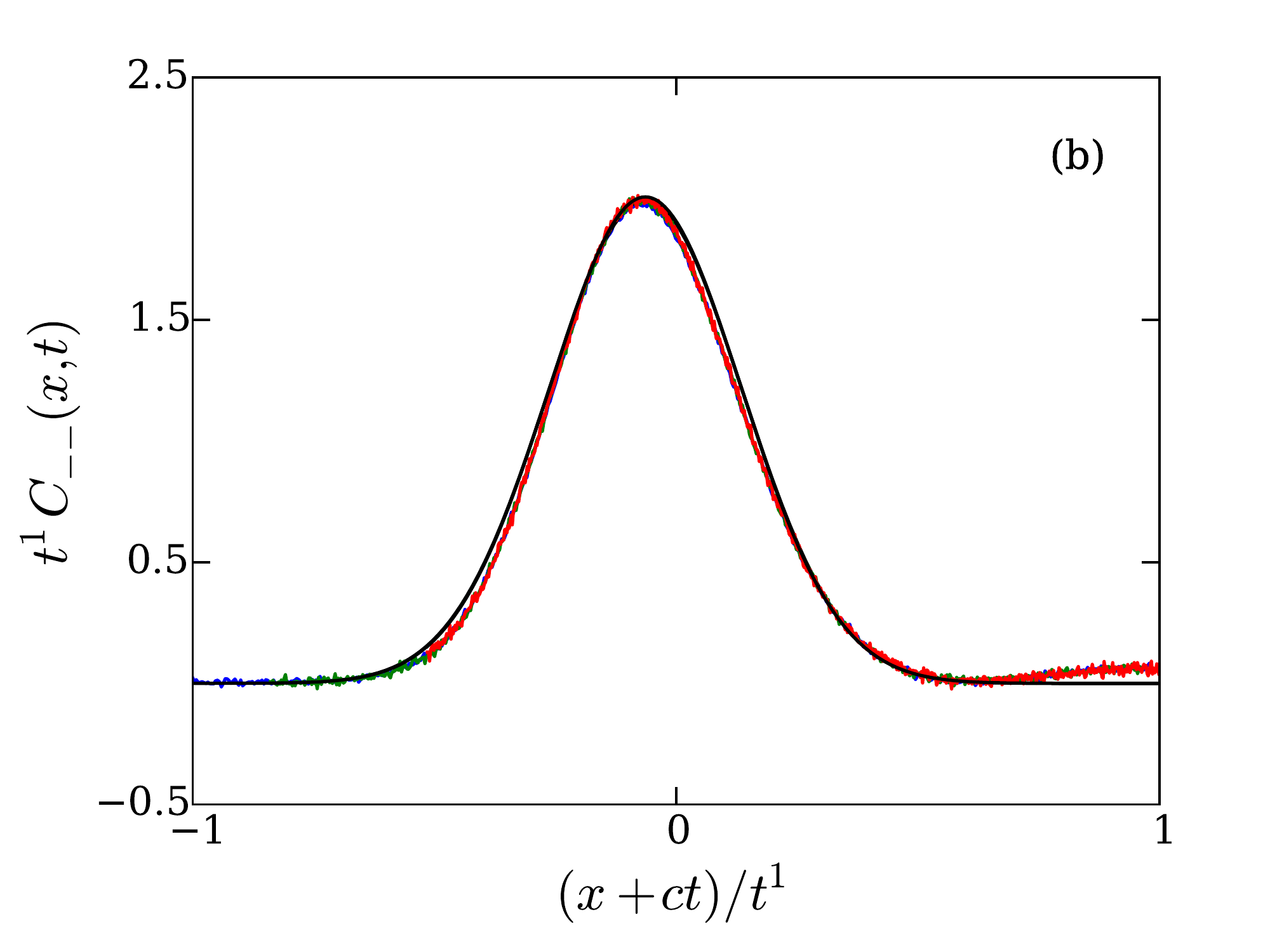}
\includegraphics[width=0.5\textwidth,height=\textheight,keepaspectratio ]{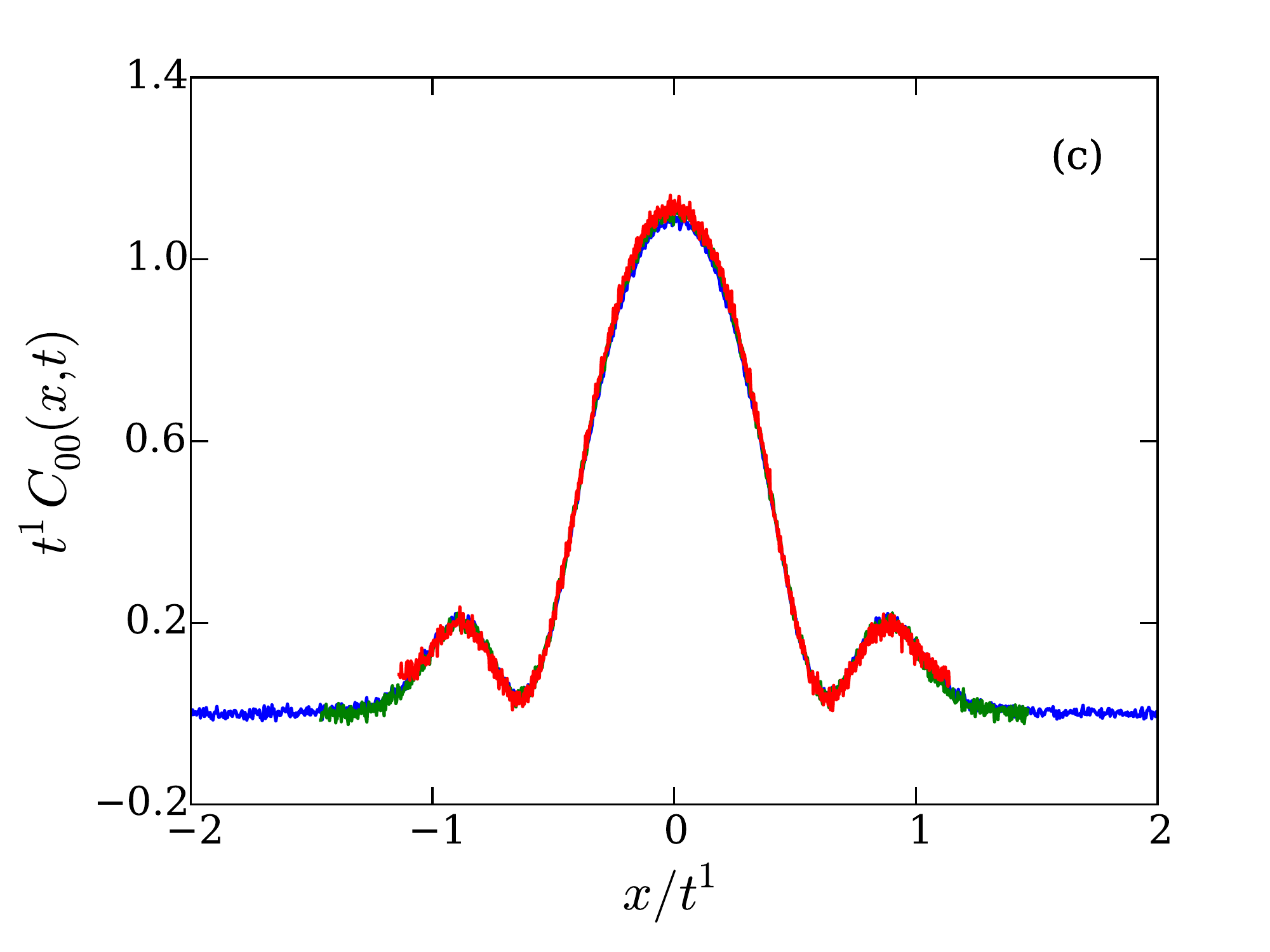}
\caption{ Normal mode representation - Case II with parameters parameters $a=1.0, b=1.0, P=1.0, T=5.0$ and $N=1024$. (a) Heat and sound modes plotted together , at times $t=200$ (red), $t=300$ (blue) and $t=450$ (green). The two sound modes move to the left and to the right with speed $c$  (vertical dashed lines indicate the position $ct$). (b) This shows the ballistic scaling of the  sound modes at the three different times. These  are now almost Gaussian (shown by black solid line with standard deviation $\sigma = 0.1982...$). (c) This shows the ballistically scaled heat mode. The amplitude of the heat and sound modes are now comparable.}
\label{T5}
\end{figure}

{\textbf {Case II:} $a=1.0, b=1.0, T=5.0, P=1.0$} --- In Fig.~\ref{T5}(a) we show the three normal modes correlations plotted together.  The speed of sound in this case is $0.6232...$. At high temperatures the dynamics is controlled by solitons, which are moving slower than their phonon counterparts.  At this temperature the phonon-soliton interaction is negligible and the sound mode is symmetric and fits well to a Gaussian with $\sigma = 0.1982$, while the heat mode has faster decay. Another feature is that at  high temperatures the diagonal correlations are at least an order of magnitude larger than the cross-correlations. In Figs.~\ref{T5}(b) and \ref{T5}(c) we show the ballistic scaling of the left moving sound mode and the heat mode.

\begin{figure}
	\includegraphics[width=0.5\textwidth,height=\textheight,keepaspectratio ]{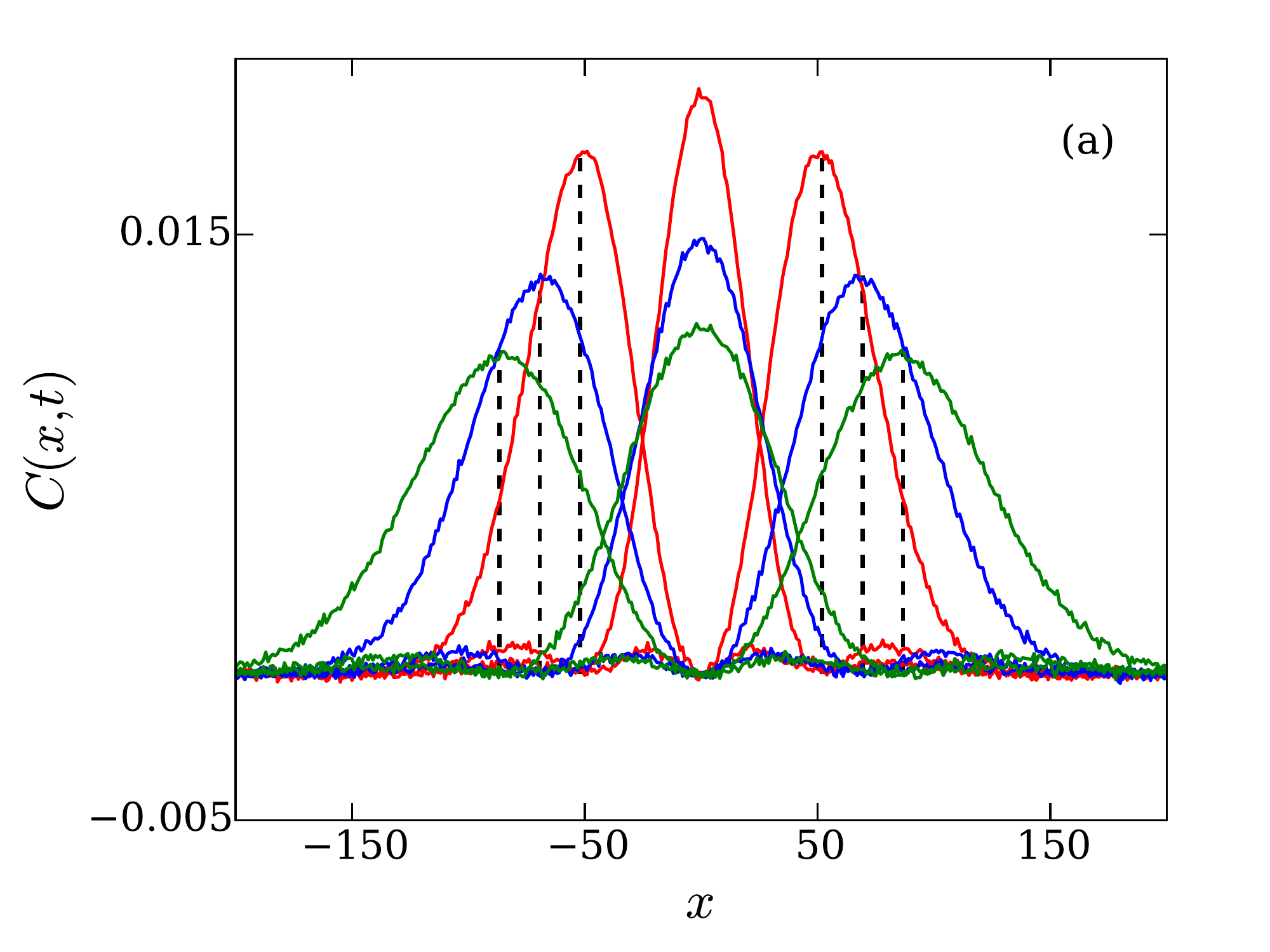}
	\includegraphics[width=0.5\textwidth,height=\textheight,keepaspectratio ]{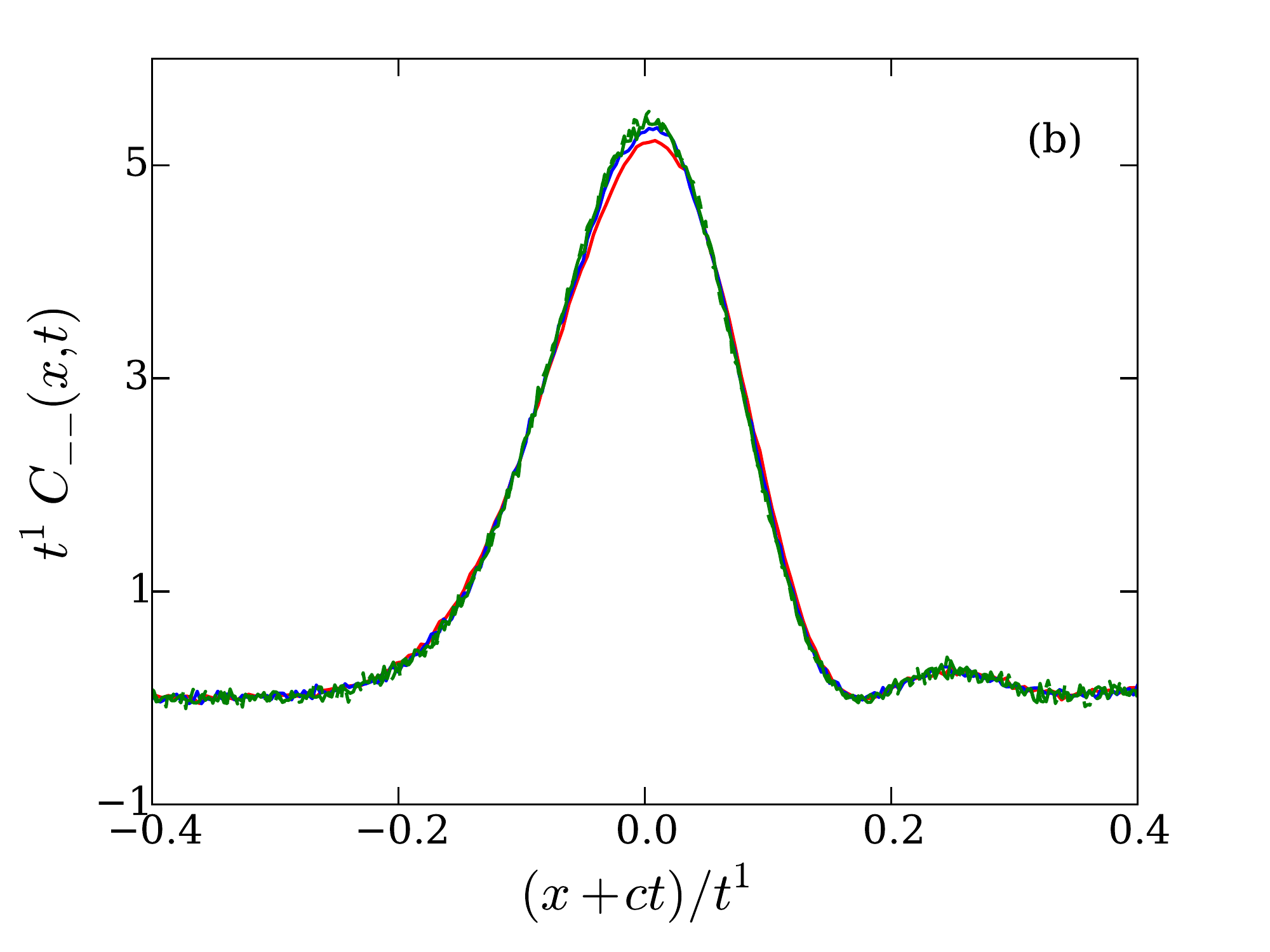}
\includegraphics[width=0.5\textwidth,height=\textheight,keepaspectratio ]{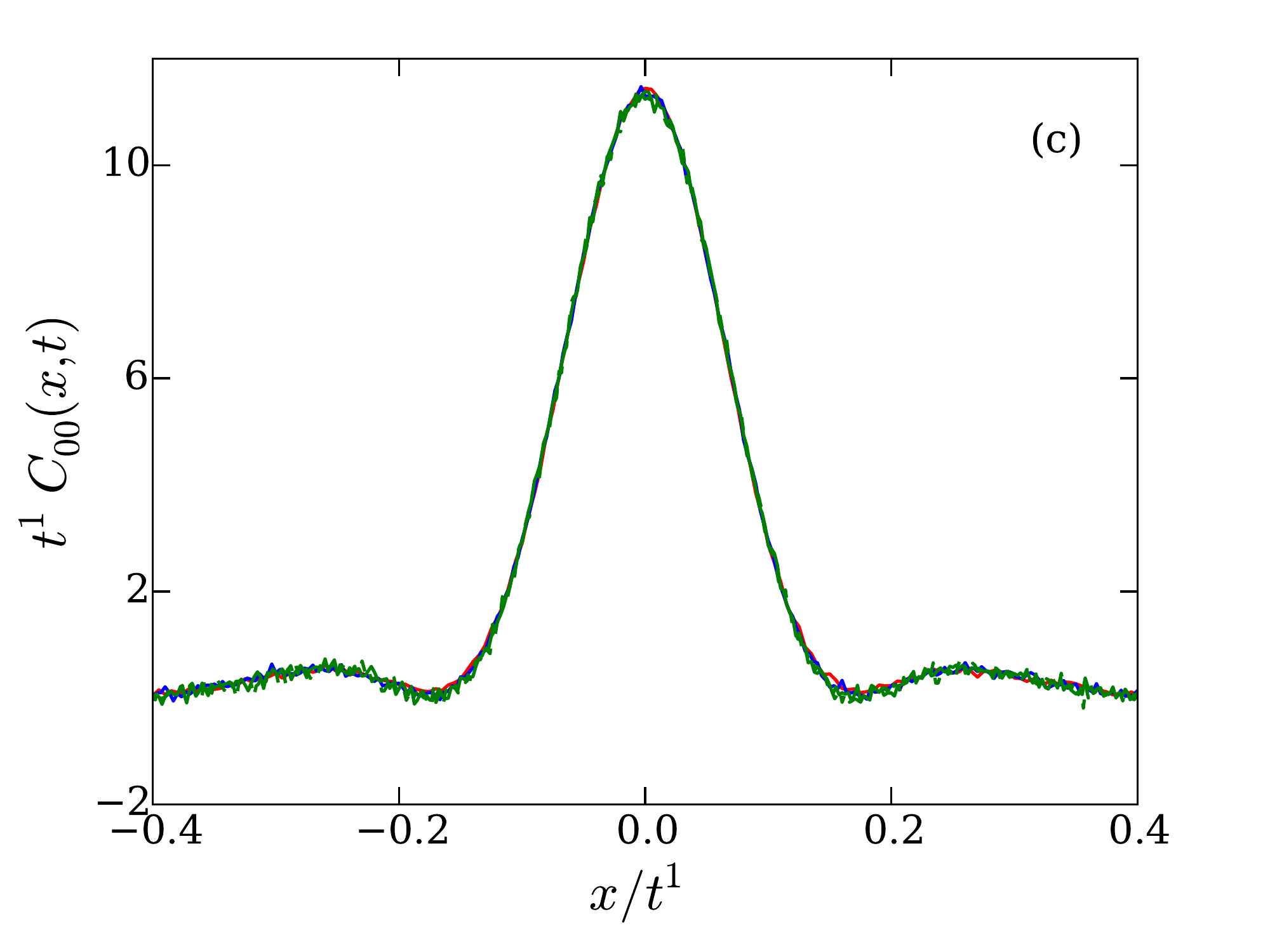}
\caption{Normal mode representation - Case III with parameters $a=0.1, b=10.0, P=0.1, T=1.0$ and  $N=1024$, corresponding to the large anharmonicity limit.  
(a) This shows the heat and the two sound modes   at times $t=300$ (red), $t=400$ (blue) and $t=500$ (green). The  distances $x=\pm ct$  are marked with vertical dashed lines. (b) This shows the ballistic scaling of the  sound modes. (c) This shows ballistic scaling of the heat mode.}
\label{HardP}

\end{figure}

{\textbf {Case III:}  $a=0.1, b=10.0, T=1.0, P=1.0$} --- In Fig.~\ref{HardP}(a) we show the three normal modes correlations plotted together, in a parameter regime corresponding to the hard particle limit. 
The speed of sound is $0.17093$.  The heat and sound modes now have single peaks but these are broad and with significant  overlap  at all times. Also note that the heat mode is larger in amplitude than the two sound modes unlike the other cases. In Figs.~\ref{HardP}(b) and \ref{HardP}(c) we show the ballistic scaling of the right moving sound mode and the heat mode.

\begin{figure}

\includegraphics[width=0.5\textwidth,height=\textheight,keepaspectratio ]{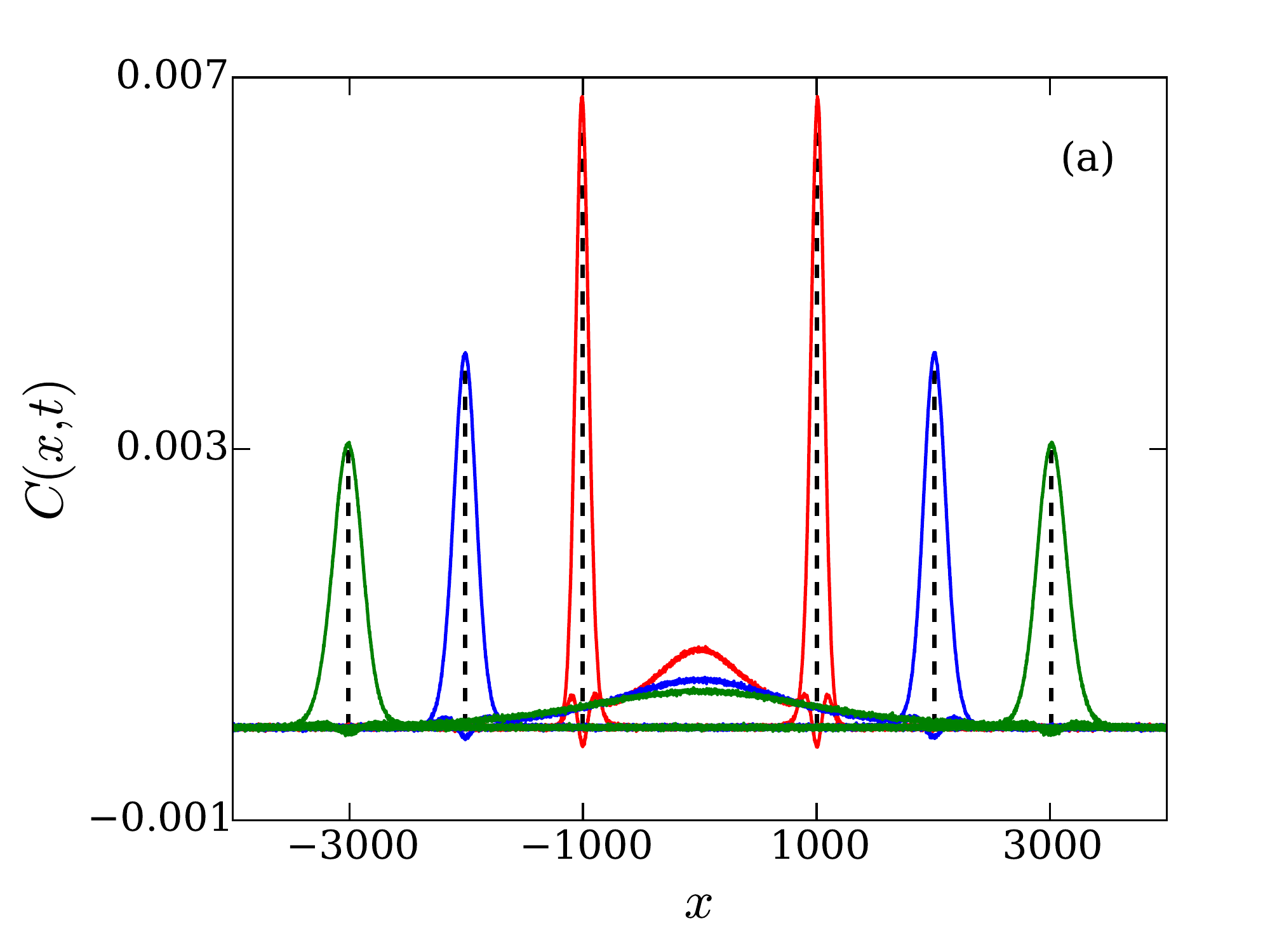}
\includegraphics[width=0.5\textwidth,height=\textheight,keepaspectratio ]{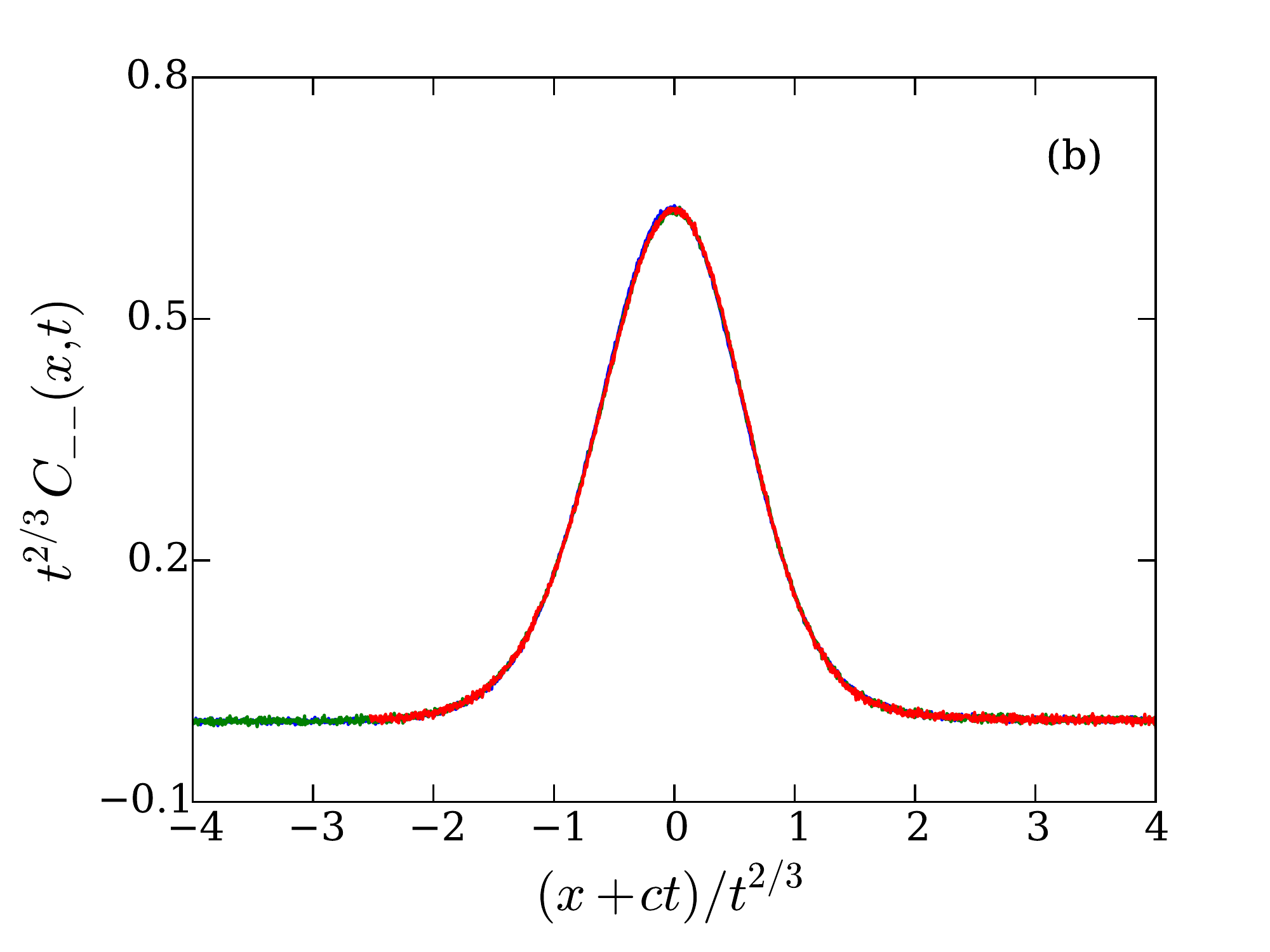}
\includegraphics[width=0.5\textwidth,height=\textheight,keepaspectratio ]{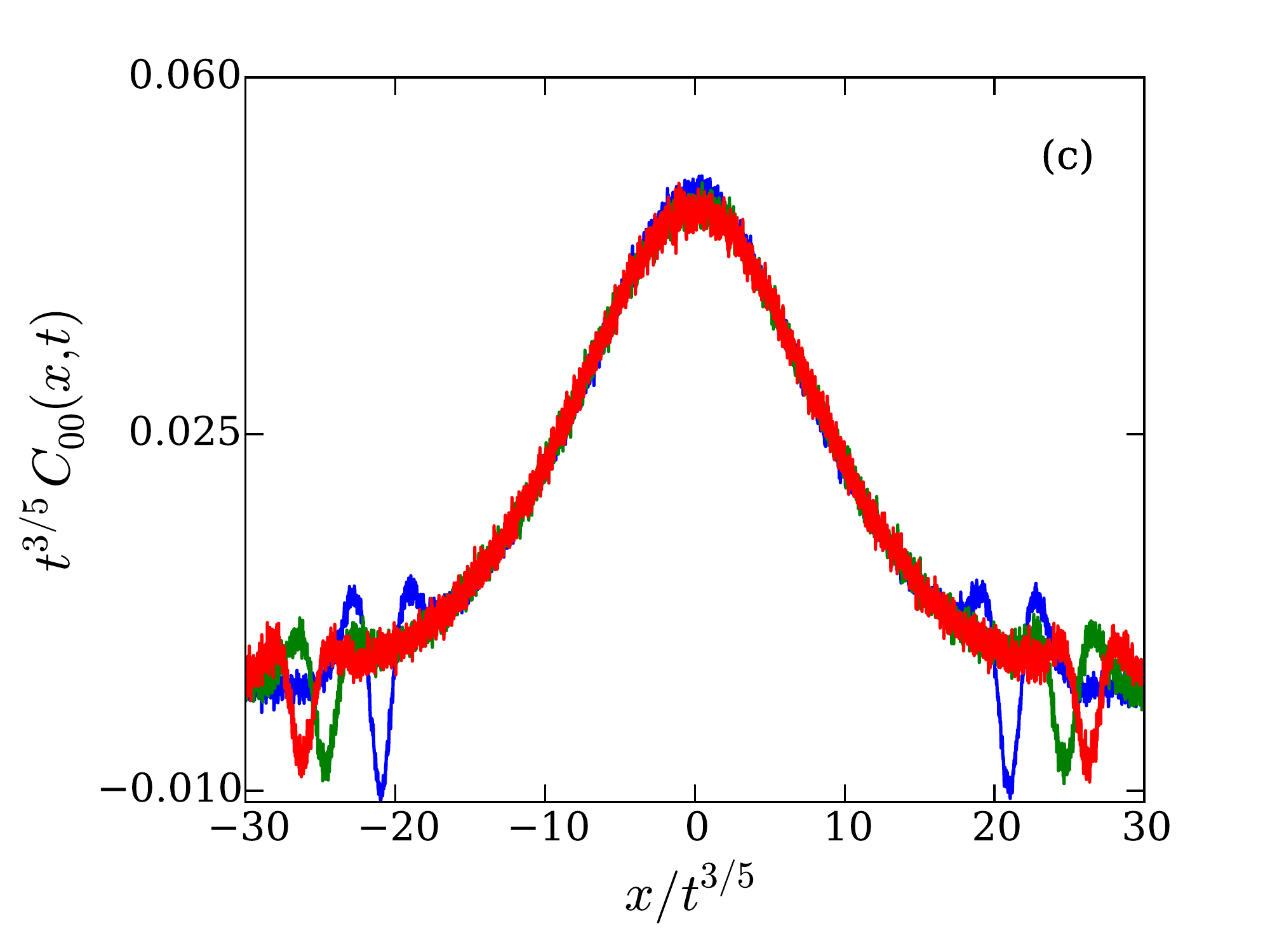}
\caption{(a) Sound and Heat modes for truncated Toda chain (Eq.~ (\ref{TruncTodaPot})) with parameters $P=0.0$, $T = 0.5$, $N = 8192$  at times $t = 2000$ (red)  $t=3000$ (blue) and $t=3500$ (green). The black dashed line show the positions $\pm ct$ and coincide with the peaks of the sound modes. 
 (b) This shows the expected KPZ-scaling of the sound modes, with 
 exponent $2/3$ as per hydrodynamics prediction. (c) This shows the heat mode scaling  with the the expected Levy exponent $3/5$.  }
\label{TruncToda}
\end{figure}
\begin{figure}
\includegraphics[width=0.55\textwidth,height=\textheight,keepaspectratio] {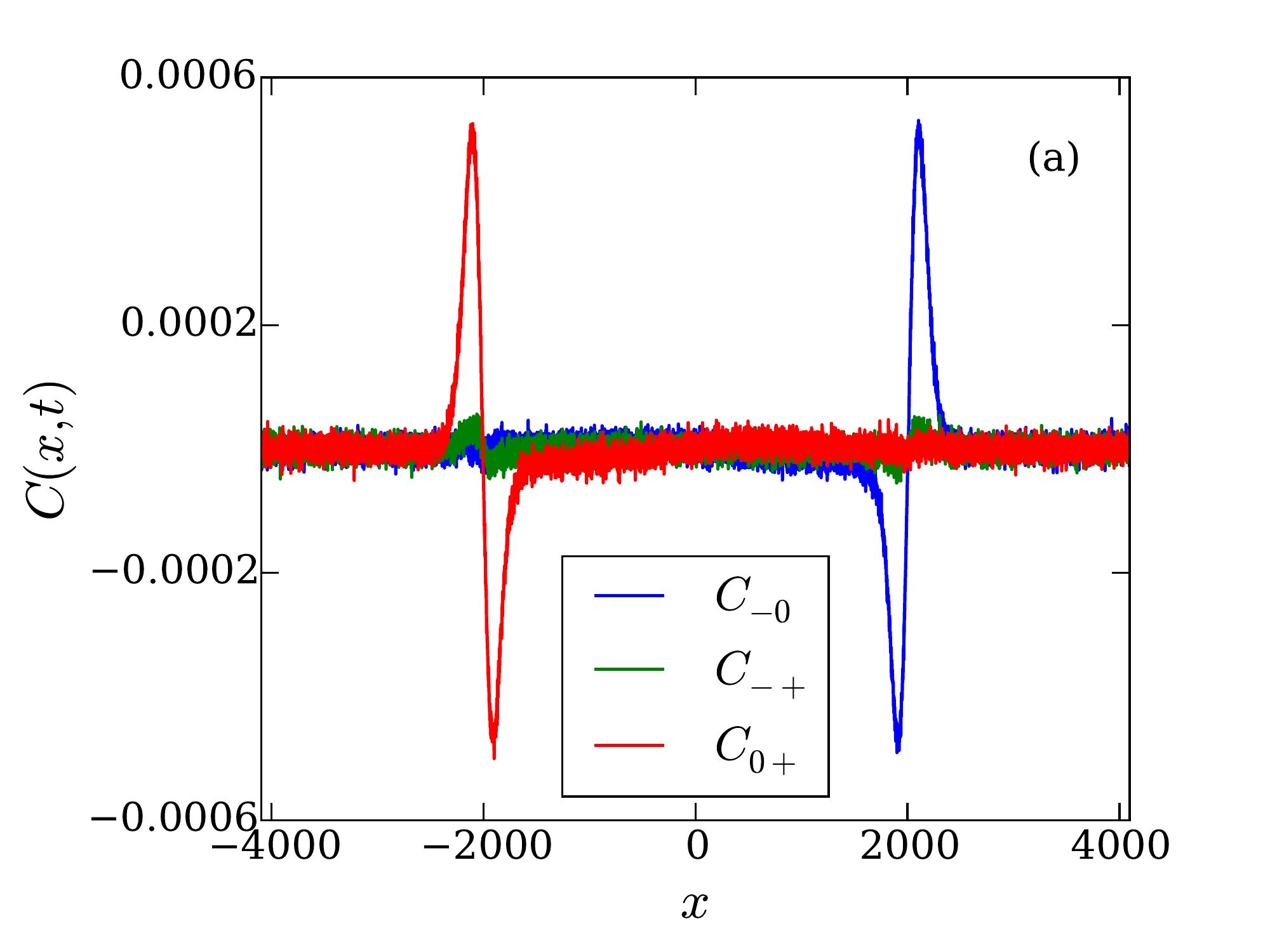}
\includegraphics[width=0.55\textwidth,height=\textheight,keepaspectratio] {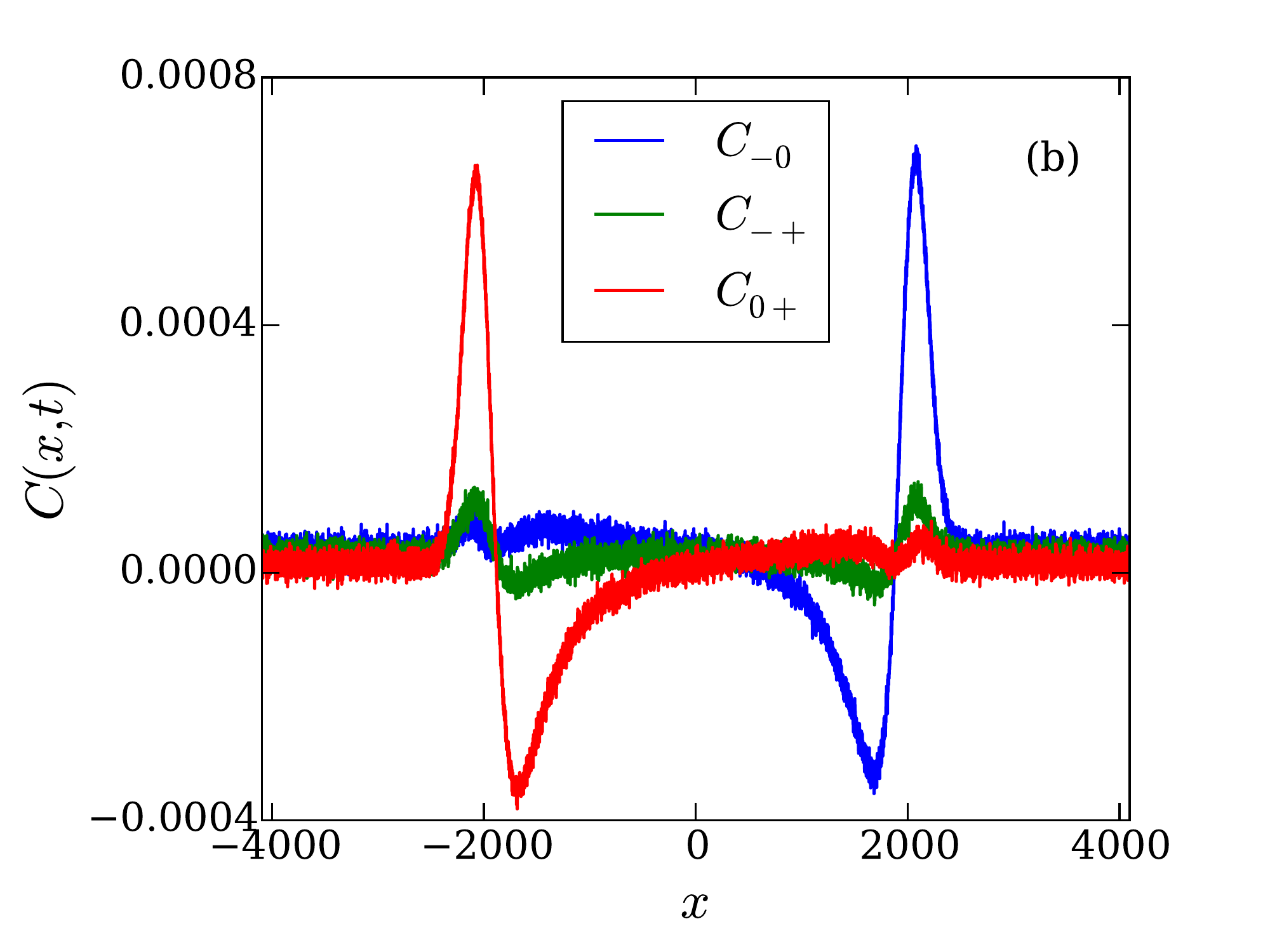}
\caption{This shows the cross correlations between the normal modes at $t = 2000$ which are smaller than their respective diagonal correlations for (a) Truncated Toda chain with parameters as given in Fig.~\ref{AllCorr}(a). (b)  Toda chain with same parameters as Fig.~\ref{AllCorr} (b). }
\label{cross_nm}
\end{figure}

Finally we show that the normal mode representation also brings out clearly the striking differences between integrable and non-integrable models.  
In Figs.~\ref{TruncToda}(a) and \ref{TruncToda}(b) we plot the normal mode correlations for the truncated Toda chain whose correlations (in usual variables) were presented in Fig.~\ref{AllCorr}(a).  We see the striking differences between these and the corresponding plots for the Toda chain in Figs.~(\ref{T1},\ref{T5},\ref{HardP}). In particular we see that for the non-integrable case, the sound modes show the KPZ scaling form $C_{++}(x,t) = f_+((x+ct)/(\lambda_s t^{2/3}))/(\lambda_s t)^{2/3}$, while the heat modes show Levy-$5/3$ scaling $C_{00}=   f_0(x/(\lambda_h t^{3/5}))/(\lambda_h t)^{3/5}$, where 
$f_{+,0}$ and $\lambda_{s,h}$ are appropriate scaling functions and scaling factors.
The cross correlation between the three normal modes in the truncated Toda lattice is shown in Fig.~ \ref{cross_nm}(a)  and for Toda chain in Fig.~ \ref{cross_nm}(b). In this case we see that for both the  Toda chain and its truncated version, the off-diagonal correlations between heat and sound modes are much 
smaller than the diagonal correlations. 
The main difference between the two cases is that in the truncated Toda chain, the modes are localized, while for integrable Toda chain they  have a broad spreading.

\section{Conclusion}
\label{conc}

We have studied the spatio-temporal equilibrium correlation functions of the fluctuations of three conserved quantities (stretch, momentum and energy) in the Toda chain. We found analytical expressions of these correlations in two different limits of the Toda chain, namely harmonic chain and hard particle gas and verified them in direct molecular dynamics simulations. The two limits can be argued to correspond to either phonon dominated dynamics or soliton dominated dynamics.

For generic parameter regimes, our numerical data shows that the Toda correlations always exhibit ballistic scaling.  We  pointed out that this form  is completely different from the correlations seen in a truncated Toda potential, which exhibits the universal scaling forms predicted by nonlinear fluctuating hydrodynamics of generic anharmonic chains. We carried out  the transformation to  normal modes following the approach of  hydrodynamics (for the three variables) and found that this  is still useful in separating the multiple  peaks seen in correlation functions of the conserved variables. Also, an explicit formula for the speed of sound is obtained. Unlike  non-integrable systems, the normal modes 
have peaks with large width (both mean position and width of the peaks scale linearly with time). 
Ballistic scaling of space-time correlations seems to be a generic feature of classical integrable systems, and proving this rigorously remains an open interesting problem. The question is also of interest in the context of integrable quantum systems.

\section{Acknowledgments}
We thank Henk van Beijeren, Sanjib Sabhapandit, Rukmini Dey and Vishal Vasan for discussions.  AD would like to thank  support   from the Indo-Israel joint research project No. 6-8/2014(IC) and from the  French Ministry of Education through the grant ANR (EDNHS)​. ​

\section{Appendix}
\subsection{Harmonic Chain correlation functions}

The Hamiltonian for the harmonic chain is given by
\be
H = \sum_{x=1}^N \frac{p^2_x}{2} + \frac{\omega^2 r^2_x}{2}~,
\ee
where $r_x = q_{x+1} - q_x$ and we assume periodic boundary conditions $r_{0}=r_N$ and $p_{N+1}=p_1$. The variables $\{r_x,p_x\}$ satisfy the equations of motion 
\bea
\partial_t{r_x} &=& p_{x+1} - p_x~,\\
\partial_t{p_x} &=& \omega^2( r_{x} - r_{x-1})~. \nn
\eea
Defining Fourier transform variables $\tilde{r}_k=\sum_{x=1}^N e^{i k x} r_x,~
\tilde{p}_k=\sum_{x=1}^N e^{i k x} p_x $,  these  satisfy the equations
\bea
\partial_t \begin{pmatrix}r_k\\p_k\end{pmatrix} &=& \hat{T}  \begin{pmatrix}r_k\\p_k\end{pmatrix} \nn \\
{\rm where}~~ \hat{T}&=& 2\iu \sin(k/2)\begin{pmatrix}0&e^{\iu k/2}\\\omega^2 e^{-\iu k/2}&0\end{pmatrix}~. \label{rpsoln} \nn
\eea
Let $\hat{S}$ be the matrix which diagonalizes $\hat{B}$, i.e, $\hat{S}^{-1} \hat{B} \hat{S} = i \Lambda$. Then the solution of the above equation is given by 
\bea
\begin{pmatrix}\tilde{r}_k(t)\\\tilde{p}_k(t)\end{pmatrix} &=& \hat{S} e^{\iu \Lambda t} \hat{S}^{-1} \begin{pmatrix}\tilde{r}_k(0)\\\tilde{p}_k(0) \end{pmatrix}\\
&=&  \begin{pmatrix}
\cos(\lambda t) & \frac{\iu e^{\iu k/2}}{\omega} \sin(\lambda t)\\
{\iu \omega e^{-\iu k/2}} \sin(\lambda t) & \cos(\lambda t) \end{pmatrix} \begin{pmatrix}\tilde{r}_k(0)\\\tilde{p}_k(0) \end{pmatrix} \nn~,
\eea
where $\lambda=2 \omega \sin (k/2)$.
The translational invariance of the problem means that the correlation matrix 
\bea
C(x,t)=  \begin{pmatrix}
\la r_x(t) r_0(0)\ra  & \la r_x(t) p_0(0)\ra \\
\la p_x(t) r_0(0)\ra &\la p_x(t) p_0(0)\ra  \end{pmatrix}
\eea
is given by
\bea
C(x,t)&=&\frac{1}{N} \sum_k \tilde{C}(k,t) e^{-\iu k x},~\label{Cxp} \\
{\rm where}~\tilde{C}(k,t)&=&\begin{pmatrix}
\la \tilde{r}_k(t) \tilde{r}_{-k}(0)\ra  & \la \tilde{r}_k(t) \tilde{p}_{-k}(0)\ra \\
\la \cx{p}_k(t) \cx{r}_{-k}(0)\ra &\la \cx{p}_k(t) \cx{p}_{-k}(0)\ra  \end{pmatrix}~.
\eea
Using the solution in Eq.~(\ref{rpsoln}) and the fact that (since the initial distribution is taken from a Gibbs ensemble with temperature $T$)
\bea
\tilde{C}(k,0)&=&\begin{pmatrix}
\la \tilde{r}_k(0) \tilde{r}_{-k}(0)\ra  & \la \tilde{r}_k(0) \tilde{p}_{-k}(0)\ra \\
\la \cx{p}_k(0) \cx{r}_{-k}(0)\ra &\la \cx{p}_k(0) \cx{p}_{-k}(0)\ra  \end{pmatrix} = \begin{pmatrix} T/\omega^2 &0 \\ 0 & T\end{pmatrix}~, \nn
\eea
we get 
\bea
\tilde{C}(k,t) 
&=& T \begin{pmatrix}
\cos(\lambda t)/\omega^2 & \frac{\iu e^{\iu k/2}}{\omega} \sin(\lambda t)\\
\frac{\iu e^{-\iu k/2}}{\omega} \sin(\lambda t) & \cos(\lambda t)
\end{pmatrix}~. \nn
\eea
Doing inverse Fourier transform gives $C(x,t)$  [Eq.~(\ref{Cxp})]. After straightforward manipulations  and going to large $N$ limit we get the following explicit correlation matrix 
\bea
C_{rr}(x,t) &=& T {\cal J}_{2|x|} (2\omega t) /\omega^2 \\
C_{rp}(x,t) &=& T (-\frac{{\cal J}_{2|x|-1} (2\omega t)}{\omega}\Theta(-x)  + \frac{{\cal J}_{2|x|+1} (2\omega t)}{\omega}\Theta(x)) \nn \\
C_{pr}(x,t) &=&  T(-\frac{{\cal J}_{2|x|+1} (2\omega t)}{\omega}\Theta(-x)  + \frac{{\cal J}_{2|x|-1} (2\omega t)}{\omega}\Theta(x)) \nn \\
C_{pp}(x,t)  &=&  T {\cal J}_{2|x|} (2\omega t) \nn
\eea
where ${\cal J}_n(z)$ is the Bessel function of first kind  and $\Theta(x)$ is the Heaviside theta function.
Since the process is Gaussian, the energy correlation is derived using expressing higher order moments in terms of two-point correlation functions.
$C_{ee}(x,t)  = [C^2_{rr}(x,t)  + C^2_{rp}(x,t)  + C^2_{pr}(x,t)  + C^2_{pp}(x,t)]/2$.

\subsection{Hard Particle Gas correlation function}
In the hard-particle limit, the particles simply exchange velocity when they collide with each other. Thus the system can  effectively be mapped to a gas of non-interacting particles, where particles exchange their identity on each collision. Indeed this mapping to the non-interacting gas was used by Jepsen \cite{JepsenJMP65} to obtain an exact 
solution for velocity-velocity autocorrelation functions in the hard-particle gas. A simpler approach was recently proposed in \cite{DharSabhapanditJSP15} to obtain the velocity-velocity autocorrelation function and we have extended this to obtain other correlations \cite{Kunduetal16}. Here we present a heuristic approach which gives the asymptotic exact results. 

Since the initial
velocities are chosen independently for each particle, the contribution to the correlation
function $\la v_r(t)v_0(0)\ra$ is non-zero only when the velocity of the $r$th particle at time $t$ is the same as that of the zero-th particle at time $0$.
The initial velocity distribution of each particle is chosen from a Maxwell distribution $f(v) =\frac{e^{- v^2/2\bar{v}^2}}{\sqrt{2 \pi}\bar{v}}$, with $\bar{v}^2=k_B T=1/\beta$. The velocity correlation function is thus approximately given by
\bea
\la v_x(t)v_0(0)\ra &=&
\int dv v^2 \delta(x-\rho vt) \frac{f(v/\bar{v})}{\bar{v}} \nn \\
&=&  \frac{\bar{v}^2}{\sigma_t} \left({\frac{x}{\sigma_t}}\right)^2 \frac{e^{-\frac{1}{2}({\frac{x}{\sigma_t}})^2}}{\sqrt{2 \pi}}~, \label {hpvv}
\eea
where  $\sigma_t = \rho\bar{v} t$.
To  compute the stretch correlations,  we note that
\begin{align}
&\la r_x(t) r_0(0) \ra  = \la [ q_{x+1}(t) -q_x(t) )(q_1(0) -q_0(0)] \ra \nn \\
&~~~~= -  \left[ \la q_{x+1}(t) q_0(0) \ra - 2 \la q_x(t) q_0(0) \ra + \la q_{x-1}(t) q_0(0) \ra \right] \nn\\
&= -\partial_x^2 \la(q_x(t)q_0(0) \ra~,
\end{align}
where we have used the translation symmetry of the chain. 
Now taking two time derivatives  gives
\begin{align}
\partial_t \la  r_x(t) r_0(0) \ra  &=  -\partial^2_x \la(v_x(t) q_0(0) \ra \nn =  -\partial^2_x \la(v_x(0)q_0(-t) \ra \nn~, \\
\partial^2 _t \la r_x(t) r_0(0) \ra &=  \partial^2_x \la(v_x(0)v_0(-t) \ra \nn 
=  -\partial^2_x \la(v_x(t)v_0(0) \ra \nn, 
\end{align}
where we used time-translation invariance.
Using this, the stretch correlation can be written in terms of velocity correlations as follows
\bea
\la  r_x(t) r_0(0) \ra  = \int_{0}^{t} dt' \int_{0}^{t'}dt''  \partial^2_x \la v_x(t)v_0(0) \ra\nn~.
\eea
This finally gives (taking the continuous $x$ limit):
\be
\la  r_x(t) r_0(0) \ra  = \frac{1}{\rho^2 \sigma_t}  \frac{e^{-\frac{1}{2}({\frac{x}{\sigma_t}})^2}}{\sqrt{2 \pi}}~.
\ee
For Energy correlation, we need to compute  
\eqa{
\la e_x(t);e_0(0) \ra &=  \la e_x(t)e_0(0) \ra - \la e_x(t) \ra \la e_0(0) \ra \nn \\
&=\frac{1}{4} \la [v^2_x(t) - \la v^2_x(0) \ra] [ v^2_0(0) -\la v^2_0(0)\ra ] \ra  \nn~.
}
A similar computation  as that leading to Eq.~(\ref{hpvv}) gives
\be
\la e_x(t);e_0(0) \ra = \frac{\bar{v}^4}{\sigma_t} \left[ \left( \frac{x}{\sigma_t}\right)^4 - 2\left( \frac{x}{\sigma_t}\right)^2 + 1 \right] f\left(\frac{x}{\sigma_t}\right)~.
\ee

\subsection{Sum Rules}
Here we outline proof's of the the sum rules mentioned in Sec.~(\ref{model}). 
The zeroth  sum rule says that for a conserved quantity, the total 
correlation of the system remain constant in time, i.e, 
\be
 \sum_{x}  C^{\alpha\beta}(x,t) = \sum_x C^{\alpha\beta}(x,0)~. \label{sumr0}
\ee
Recall that we are interested in correlations of the fluctuations around equilibrium values $u_\alpha(x,t) =  I_\alpha(x,t) - \la I_\alpha \ra$. 
Let us also define the current fluctuations as $\Delta j_\alpha(x,t) = j_\alpha(x,t)- \la j_\alpha \ra$ and the total current $J^\alpha(t)=\sum_{x} j^\alpha(x,t)$. From the equations of motion we get
\be
\partial_t u^{\alpha}(x,t) =   \Delta j^\alpha_{x-1}(t) -  \Delta j^\alpha_x(t)~. \nn 
\ee
Multiplying both sides by $u^{\beta}(0,0)$ and averaging over the initial equilibrium distribution gives 
\eqa{
\partial_t C^{\alpha\beta}(x,t)  &=   \la  \Delta j^\alpha_{x-1}(t)u^{\beta}(0,0) \ra - \la  \Delta j^\alpha_x(t)u^{\beta}(0,0) \ra \nn \\
&=  \la   \Delta  j^\alpha_{0}(0)u^\beta_{1-x}(-t) \ra - \la   \Delta j^\alpha_0(0) u^\beta_{-x}(-t)) \ra~, \label{Cdt}
}
where we used space and time-translational invariance. Summing over all sites we then get
\eqa{
\frac{d}{dt} \sum_x C^{\alpha\beta}(x,t) &= \sum_x [ \la\Delta j^\alpha_{0}(0)u^\beta_{x-1}(-t) \ra -\la\Delta  j^\alpha_0(0) u^\beta_x(-t) \ra] \nn 
}
which vanishes, since $\sum u^\beta_x$ is a conserved quantity. Hence the result in Eq.~(\ref{sumr0}) follows.

The other sum rules are on  the  moments of spatial correlation functions 
of conserved quantities. The first and second sum rules respectively state 
\eqa{
\frac{d}{dt} \sum_x  x C^{\alpha\beta}(x,t) &= \sum_{x=-N/2}^{N/2-1} \la \Delta j^\alpha(x,0) u^\beta(0,0) \ra \nn \\ & -N \la j^\alpha(-N/2,t) ^{\beta}(0,0) \ra ~, \nn \\
& = \la J^\alpha u^\beta \ra ~~~ (N \to \infty) \label{sumr1}  \\
\frac{d^2}{dt^2} \sum_x  x^2 C^{\alpha\beta}(x,t) &= 2 \sum_{-N/2}^{N/2-1} C^{\alpha\beta}_j(x,t)  \nn \\&
+ N \left[ C^{\alpha\beta}_{j}(-\frac{N}{2},t) - C^{\alpha\beta}_{j}(\frac{N}{2}-1,t)\right] \nn \\
&= 2  \sum_x C^{\alpha\beta}_j(x,t)  \label{sumr2} ~~~~(N\to \infty)
} 
where $C^{\alpha \beta}_j(x,t) = \la \Delta j^\alpha (x,t) \Delta j^\beta(0,0)\ra$ and we note that Eq.~(\ref{sumr1}) only involves an equilibrium equal time correlation. 

The proof starts by following steps as those for  Eq.~(\ref{Cdt}) to get
\be
\partial_t C^{\alpha\beta}(x,t)= \la \Delta j^\alpha_{x-1}(0)u^\beta_0(-t) \ra - \la \Delta j^\alpha_x(0) u^\beta_0(-t) \ra~. \nn 
\ee
Multiplying the above equation by $x$, summing over all $x$, and after simplifications using the fact that $\sum_x u^{\beta}(x,t) = const$ gives Eq.~(\ref{sumr1}).
Taking another time derivative, and on using the continuity equations we get
\eqa{
\frac{d^2}{dt^2} C^{\alpha\beta}(x,t) &= -[ \la \Delta j^{\alpha}_{x-1}(0) [\Delta j^{\beta}_{-1}(-t) - \Delta j^{\beta}_0(-t)] \ra \nn\\ & - \la \Delta j^{\alpha}_x(0) [\Delta j^{\beta}_{-1}(-t) - \Delta j^{\beta}_0(-t)] \ra ]\nn \\
 &= \left[ C^{\alpha\beta}_j(x+1,t) - 2 C^{\alpha\beta}_j(x,t)+ C^{\alpha\beta}_j(x-1,t)\right]~. \nn
}
Multiplying the above equation by $x^2$, summing over all $x$, and after simplifications using the first sum-rule $\sum_x C^{\alpha\beta}(x,t) = const$ gives Eq.~(\ref{sumr2}).


\end{document}